\documentclass[aps,twocolumn,floats,prd,nofootinbib,superscriptaddress,10pt]{revtex4-1}

\usepackage{graphicx,amsmath,amssymb}
\usepackage{hyperref}
\usepackage{sidecap}
\usepackage{xcolor}
\sidecaptionvpos{figure}{m}
\usepackage[hang,flushmargin, norule]{footmisc} 
\usepackage{pifont}

\hypersetup{colorlinks=true, linkcolor=red, filecolor=magenta, urlcolor=blue}
\AtBeginDocument{
    \newwrite\bibnotes
    \def\bibnotesext{Notes.bib}
    \immediate\openout\bibnotes=\jobname\bibnotesext
    \immediate\write\bibnotes{@CONTROL{REVTEX41Control}}
    \immediate\write\bibnotes{@CONTROL{
    apsrev41Control,author="08",editor="1",pages="1",title="0",year="1"}}
     \if@filesw
     \immediate\write\@auxout{\string\citation{apsrev41Control}}
    \fi
}

\newcommand{\customfootnotetext}[2]{{
\renewcommand{\thefootnote}{#1}
\footnotetext[0]{#2}}}

\begin{document}

\title{Does it matter? A more careful treatment of density fluctuations in 21-cm simulations}

\author{Jordan Flitter}
\email{E-mail: jordanf@post.bgu.ac.il}
\author{Sarah Libanore}
\author{Ely D.\ Kovetz}
\affiliation{Physics Department, Ben-Gurion University of the Negev, Beer-Sheva 84105, Israel}

\begin{abstract}

The cosmological 21-cm signal is sourced from hyperfine transitions in neutral hydrogen atoms. Yet, although the abundance of hydrogen atoms follows the baryon density field, semi-numerical codes that simulate the 21-cm signal simplify their treatment as if all the matter in the Universe was 
in the form of collisionless cold dark matter (CDM). This is usually done by evolving the density field via a scale-independent growth factor (SIGF). In this work, we separate the baryons from CDM and evolve the two species with a proper scale-dependent growth factor (SDGF). By incorporating the SDGF in the {\tt 21cmFirstCLASS} code, we demonstrate the effect that baryons and CDM have on the 21-cm signal at the linear dark ages epoch and the subsequent non-linear epochs of cosmic dawn and reionization. Our analysis shows that the baryonic nature of hydrogen cannot be ignored during the dark ages, and that non-linear effects in density-field evolution must be accounted for after stars have formed. Furthermore, we discuss how the 21-cm signal is modified at lower redshifts, where ground-based 21-cm interferometers are mostly sensitive, due to the choice of working with  either the “linear” or “non-linear” matter over-density (that is, the over-density as computed from linear perturbation theory, and non-linear perturbation theory, respectively) in the extended Press-Schechter formalism. Our code is publicly available at \href{https://github.com/jordanflitter/21cmFirstCLASS}{https://github.com/jordanflitter/21cmFirstCLASS}.

\end{abstract}

\maketitle

\section{Introduction}
\vspace*{-.2cm}

Between recombination and reionization, the Universe was filled with neutral hydrogen atoms. Hyperfine energy de-excitations in these atoms, following 
interactions with other particles, led to the emission of 21-cm wavelength photons. This is the 21-cm signal in a nutshell~\cite{Madau:1996cs, Barkana:2000fd, Loeb:2003ya, Bharadwaj:2004it, Furlanetto:2006jb, Pritchard:2011xb, Bera:2022vhw, Shaw:2022fre}. This exciting signal has the potential to expand our knowledge on the formation mechanism of the first galaxies~\cite{Barkana:2004vb, Furlanetto:2006tf, Park:2018ljd, Munoz:2019hjh, Park:2019aul, Greig:2019tcg, Thyagarajan:2020kua, Qin:2020xyh, Qin:2020pdx, Reis:2020hrw, Ma:2023oko}, the thermal history of the intergalactic medium (IGM)~\cite{Chen:2003gc, Mesinger:2013nua, Ghara:2019kir, Greig:2020hty, Mittal:2020kjs, Ghara:2021fqo, Maity:2021mhr, HERA:2021noe, HERA:2022wmy, Lazare:2023jkg}, the epoch of reionization~\cite{Lidz:2008ry, Wiersma:2012ek, Jensen:2013fha, Meerburg:2013dua, Greig:2015qca, Greig:2017jdj, Greig:2020suk, Ghara:2020syx, Watkinson:2021ctc, Rahimi:2021wom, Qin:2021gkn, Saxena:2023tue, Giri:2024xet, Ghara:2024xri}, physics beyond the standard model~\cite{Tashiro:2006uv, Schleicher:2008hc, Shiraishi:2014fka, Chluba:2015lpa, Villaescusa-Navarro:2015cca, Oyama:2015gma, Pal:2016icc, Kovetz:2018zes, Caputo:2020avy, Natwariya:2022xlv, Kunze:2018cnn, Minoda:2022nso, Shmueli:2023box, Kunze:2023pfj, Cruz:2023rmo, Libanore:2023oxf, Bhaumik:2024efz, Lee:2023uxu, Dhuria:2024zwh, Plombat:2024kla, Adi:2024ebl, Libanore:2025ack}, and possibly shed some light on the nature of dark matter~\cite{Tashiro:2014tsa, Evoli:2014pva, Munoz:2015bca, Lopez-Honorez:2016sur, DAmico:2018sxd, Cheung:2018vww, Mitridate:2018iag, Lidz:2018fqo, Nebrin:2018vqt, Fialkov:2018xre, Barkana:2018lgd, Clark:2018ghm, Kovetz:2018zan, Berlin:2018sjs, Barkana:2018qrx, Slatyer:2018aqg, Munoz:2018pzp, Liu:2018uzy, Munoz:2018jwq, Creque-Sarbinowski:2019mcm, Liu:2019knx, Hotinli:2021vxg, Jones:2021mrs, Halder:2021uoa, Mittal:2021egv, Sarkar:2022dvl, Barkana:2022hko, Giri:2022nxq, Flitter:2022pzf, Lazare:2024uvj, Driskell:2022pax, Dey:2022ini, Short:2022bmm, Mondal:2023bxb, Facchinetti:2023slb, Qin:2023kkk, Sun:2023acy, Vanzan:2023gui, Cang:2023bnl, Shao:2023agv, Katz:2024ayw, Lopez-Honorez:2024ant}.

For these reasons, worldwide efforts are being made to measure the 21-cm signal by various collaborations and experiments. Some of these experiments are designed to measure the ``global" (sky-averaged) 21-cm signal, like the Experiment to Detect the Global reionization Signature (EDGES)~\cite{Monsalve:2019baw},  Shaped Antenna measurement of the background RAdio Spectrum (SARAS)~\cite{2021arXiv210401756N},  Large-Aperture Experiment to Detect the Dark Ages (LEDA)~\cite{2018MNRAS.478.4193P}, the Radio Experiment for the Analysis of Cosmic Hydrogen (REACH)~\cite{deLeraAcedo:2022kiu} and Probing Radio Intensity at high-Z from Marion (PRizM)~\cite{2019JAI.....850004P}. Other experiments, like the Hydrogen Epoch of Reionization Array (HERA)~\cite{DeBoer:2016tnn}, Low Frequency Array (LOFAR)~\cite{2013A&A...556A...2V}, Murchison Widefield Array (MWA)~\cite{9f7294777bdb407db9646a284ffeae0e}, the Giant Metrewave Radio Telescope (GMRT)~\cite{1991CSci...60...95S}, the Precision Array for Probing the Epoch of Reionization (PAPER)~\cite{Kolopanis:2019vbl} and Square Kilometre Array (SKA)~\cite{2015aska.confE.174B} have been dedicated to measuring the spatial fluctuations in the signal using ground based interferometers. All the aforementioned experiments are limited to detecting the signal at redshifts $z\lesssim 30$ (namely, high frequencies), as the Earth ionosphere distorts incoming low-frequency radiation that is sourced at very high redshifts. Several interesting proposals are considered nowadays to overcome this obstacle by deploying radio receivers on the far side of the Moon, or in orbit around it or the Earth~\cite{Jester:2009dw, Burns:2021pkx, Furlanetto:2019jso, Silk:2020bsr, Burns:2021ndk, Burns:2021pkx, Goel:2022jgw, Shi:2022zdx, Bertone:2023ojo, Fialkov:2023yda, Mondal:2023xjx}.

The growing interest in the 21-cm signal and its analysis motivated the development of a plethora of dedicated codes, designed to estimate the signal and its spatial fluctuations as a function of redshift. After recombination, prior to star formation, the fluctuations in the signal are tiny and linear perturbation theory can be applied in that regime. For example, {\tt CAMB}~\cite{Lewis:2007kz} and {\tt PowerSpectrum21cm}~\cite{Munoz:2015eqa} exploit the linearity of the perturbations in this ``dark ages" epoch, see also Ref.~\cite{Naoz:2005pd}. After cosmic dawn, once galaxies and stars have formed, the fluctuations in the signal are amplified considerably, and the validity of linear perturbation theory breaks. Different techniques have been adapted in the literature to simulate the signal in the non-linear regime of cosmic dawn and reionization. This includes full radiative transfer simulations ({\tt CodA}~\cite{Ocvirk:2015xzu, Ocvirk:2018pqh, Lewis:2022kwf}, {\tt 21SSD}~\cite{Semelin:2017xgv}, {\tt THESAN}~\cite{Kannan:2021xoz}), one-dimensional radiative transfer methods ({\tt BEoRN}~\cite{Schaeffer:2023rsy}, {\tt Bears}~\cite{Thomas:2008uq}, {\tt Grizzly}~\cite{Ghara:2017vby}), hydrodynamic-radiative-transfer frameworks ({\tt Licorice}~\cite{Semelin:2007rk, Meriot:2023usy}), and ray-tracing algorithms ({\tt C$^2$-ray}~\cite{Mellema:2005ht, Hirling:2023cjx}, {\tt CRASH}~\cite{Maselli:2003ij}).

Semi-numerical codes, whose runtime is shorter by orders of magnitude, are also found in the literature. Among these, {\tt 21cmFAST}~\cite{Mesinger:2010ne, Munoz:2021psm} is perhaps the most popular one, but see also {\tt simfast21}~\cite{Santos:2009zk}, {\tt CIFOG}~\cite{Hutter:2018qxa} and {\tt 21cmSPACE}~\cite{Visbal:2012aw, Fialkov:2012su, Fialkov:2013uwm, Fialkov:2014kta, Fialkov:2014wka, Fialkov:2015fua, Cohen:2015qta, Fialkov:2016zyq, Fialkov:2019vnb, Reis:2020arr, Reis:2021nqf, Reis:2021sqh, Magg:2021jyc, Gessey-Jones:2022njt, Sikder:2023ysk, Gessey-Jones:2023amq, Pochinda:2023uom}, though the latter is not publicly available as of yet. Recently, using output from {\tt 21cmFAST}, and assuming linear evolution of the density field, a fully analytic code named {\tt Zeus21}~\cite{Munoz:2023kkg, Cruz:2024fsv} was calibrated at the cosmic dawn era, allowing the code to compute the 21-cm power spectrum in mere seconds.\footnote{{\tt ARES}~\cite{Mirocha:2014faa, 2017MNRAS.464.1365M} and {\tt ECHO21}~\cite{Mittal:2025les} are also speedy codes, but unlike {\tt Zeus21} and {\tt HMreio}, they only compute the global 21-cm signal.} Another analytic code, named {\tt HMreio}~\cite{Schneider:2023ciq}, also calibrated to {\tt 21cmFAST}, focuses on the epoch of reionization. All these codes use a single quantity $\delta$ to denote over-density, and thus implicitly assume $\delta_b=\delta_c=\delta_m$, namely that  baryons (``$b$") and cold dark matter (CDM, ``$c$") behave alike (``$m$" stands for all of matter).

In this paper, we relax this assumption and explore the cosmological implications of distinguishing $\delta_b$ from $\delta_c$ in the context of the 21-cm signal. To do so, we rely on the newly developed tool {\tt 21cmFirstCLASS}~\cite{Flitter:2023mjj, Flitter:2023rzv}. We find that at the dark ages the approximation $\delta_b\approx\delta_c$ fails when high order statistics (e.g. the power spectrum) are considered, and $\delta_b$ must be evolved in a scale-dependent manner in that epoch. We also demonstrate, using scale-dependent Lagrangian perturbation theory (LPT), the need to go beyond linear perturbation theory for the baryon over-density field at low redshifts, after cosmic dawn was set. Finally, we present how the signal is modified when considering either the ``linear" or  ``non-linear" matter over-density $\delta_m$ in the context of the extended Press-Schechter formalism. The modifications made in {\tt 21cmFirstCLASS} for this work now make it the only public code in the literature that can consistently simulate the fluctuations in the 21-cm signal at all cosmological epochs, from dark ages until reionization, while simultaneously studying the cosmic microwave background (CMB) anisotropies.

This paper is organized as follows. Sec.~\ref{sec: Theory} provides the theoretical background for the 21-cm signal and the matter fields that govern its shape. In Sec.~\ref{sec: Methods} we give a short description of the main tool used in our analysis, {\tt 21cmFirstCLASS}, and the modifications made for this work. We show our results in Sec.~\ref{sec: Results}, discuss their significance in Sec.~\ref{sec: Discussion}, and conclude in Sec.~\ref{sec: Conclusions}. In addition, this paper contains several appendices. In Appendix \ref{sec: Boltzmann equation for the spin temperature}, we study the Boltzmann equation for the spin temperature. In Appendix \ref{sec: Scale-dependent matter evolution}, we study the sensitivity of the 21-cm signal to the precise evolution of CDM and the total matter density fields. We derive the equations for second-order scale-dependent LPT in Appendix \ref{sec: Scale-dependent Lagrangian perturbation theory}. Finally, in Appendix \ref{sec: Temperature fluctuations in the dark ages}, we verify the output of {\tt 21cmFirstCLASS}, by comparing the temperature fluctuations at the dark ages with the analytical prediction from linear perturbation theory.

Throughout this work, we adopt cosmological parameters from Planck18 best-fit values~\cite{Planck:2018vyg}, without including data from baryon acoustic oscillations (BAO), namely a Hubble constant of $h=0.6736$, a total matter and baryon density parameters of $\Omega_m=0.3153$ and $\Omega_b=0.0493$, a primordial curvature power spectrum with amplitude $A_s=2.1\times10^{-9}$ and spectral index $n_s=0.9649$, and an optical depth to reionization\footnote{While {\tt 21cmFirstCLASS} is capable of computing $\tau_\mathrm{reio}$ from the simulation in a self-consistent manner (see Refs.~\cite{Shmueli:2023box, Montero-Camacho:2024dzs}), the exact value of $\tau_\mathrm{reio}$ is not crucial for this work as it affects only the CMB power spectrum.} of $\tau_\mathrm{reio}=0.0544$. We also conservatively assume 
two massless neutrinos with a third massive neutrino of mass $0.06\,\mathrm{eV}$. For the astrophysical parameters, we 
work with the fiducial values of EOS2021 (see Table 1 in Ref.~\cite{Munoz:2021psm}), but we assume $L_X/\mathrm{SFR}=10^{40}\,\mathrm{erg\cdot yr\cdot sec^{-1}}\cdot M_\odot^{-1}$ for both popII and popIII stars. All of our {\tt 21cmFirstCLASS} simulations are realized with a box of size $400\,\mathrm{Mpc}$ and a simulation cell of $2\,\mathrm{Mpc}$. All the formulae in this paper are expressed in the CGS unit system. To reduce clutter, we often do not write explicitly the independent arguments of the quantities in our equations (e.g. position, redshift, wavenumber, etc) and they should be inferred from the context.

\section{Theory}\label{sec: Theory}

In this section, we summarize the main theoretical components needed to consistently model the evolution of the 21-cm signal during the dark ages and cosmic dawn. All of them, have been implemented in the public code {\tt 21cmFirstCLASS}, as described in Sec.~\ref{sec: Methods}.

\subsection{The 21-cm signal}\label{subsec: The 21-cm signal}

The quantity of most interest in the study of the 21-cm signal is the brightness temperature relative to the CMB, defined as
\begin{equation}\label{eq: 1}
T_{21}=\frac{T_s-T_\gamma}{1+z}\left(1-\mathrm{e}^{-\tau_{21}}\right).
\end{equation}
Here, $T_\gamma$ is the characteristic temperature of the radio background, which we conservatively assume to follow the CMB temperature, $T_\gamma\propto\left(1+z\right)$. We consider this background to be 
homogeneous in space on the scales of interest for 21-cm tomography, $T_\gamma=\bar T_\gamma$ (see however studies that included background sources other than the CMB~\cite{Feng:2018rje, Fialkov:2019vnb, Reis:2020arr, Sikder:2023ysk}). The 21-cm optical depth $\tau_{21}$ is given by
\begin{equation}\label{eq: 2}
\tau_{21}=\frac{3\hbar A_{10}c\lambda_{21}^2n_\mathrm{HI}}{16Hk_BT_s}\left(1+\frac{1}{aH}\frac{d\left(\hat{\mathbf n}\cdot\mathbf v_b\right)}{d\left(\hat{\mathbf n}\cdot\mathbf x\right)}\right)^{-1},
\end{equation}
where $\hbar$ is the reduced Planck constant, $A_{10}\approx2.85\times10^{-15}\,\mathrm{sec}^{-1}$ is the Einstein coefficient for spontaneous transition from the excited hyperfine level in hydrogen to the ground level, $a$ is the scale factor, $H\equiv\dot a/a$ is the Hubble parameter, $c$ is the speed of light, $\lambda_{21}\approx\,21\,\mathrm{cm}$ is the photon wavelength that is associated with hyperfine transitions, $k_B$ is Boltzmann constant, $n_\mathrm{HI}$ is the number density of neutral hydrogen atoms, $\mathbf v_b$ is the proper baryon peculiar velocity, and $\hat{\mathbf n}$ is a unit vector along the line-of-sight.

The quantity $T_s$ that appears in Eqs.~\eqref{eq: 1}-\eqref{eq: 2} is the spin temperature. This quantity is defined as $n_1/n_0\equiv3\,\mathrm{exp}\left(-T_\star/T_s\right)$, where $n_1$ ($n_0$) denotes the number density of hydrogen atoms in the hyperfine triplet (singlet) state and $T_\star\equiv\hbar c/\left(k_B\lambda_{21}\right)\approx68.2\,\mathrm{mK}$. In theory, the solution for $T_s$ has to be obtained by solving the Boltzmann equation for $n_1$ and $n_0$. Yet, we show in Appendix \ref{sec: Boltzmann equation for the spin temperature} that the solution for this equation is very well approximated by the well known expression for $T_s$ in equilibrium, even in the dark ages where $H\gg A_{10}$. This equilibrium expression reads
\begin{equation}\label{eq: 3}
T_s^{-1}=\frac{x_\mathrm{CMB}T_\gamma^{-1}+x_\mathrm{coll}T_k^{-1}+\tilde x_\alpha T_\alpha^{-1}}{x_\mathrm{CMB}+x_\mathrm{coll}+\tilde x_\alpha},
\end{equation}
where $T_k$ is the gas/baryon kinetic temperature, $T_\alpha\approx T_k$ is the color temperature that is associated with Ly$\alpha$ photons, and $x_\mathrm{CMB}\equiv\left(1-\mathrm{e}^\mathrm{-\tau_{21}}\right)/\tau_{21}\approx 1$~\cite{Venumadhav:2018uwn}, $x_\mathrm{coll}$~\cite{Furlanetto:2006jb} and $\tilde x_\alpha$~\cite{Hirata:2005mz} are the radiation, collisional and Ly$\alpha$ coupling coefficients. The latter is proportional to the Ly$\alpha$ flux originating from stars, $J_\alpha$, and thereby is irrelevant during the time of the dark ages.

In order to compute $T_s$ one has to solve the following differential equation for $T_k$ that arises from the first law of thermodynamics,
\begin{eqnarray}\label{eq: 4}
\nonumber\frac{dT_k}{dt}=&&-2HT_k+\frac{2}{3}\frac{T_k}{1+\delta_b}\frac{d\delta_b}{dt}-\frac{T_k}{1+x_e}\frac{dx_e}{dt}
\\&&+\left.\frac{dT_k}{dt}\right|_\mathrm{CMB}+\left.\frac{dT_k}{dt}\right|_X,
\end{eqnarray}
where $\delta_b$ is the baryon energy over-density and $x_e\equiv n_e/\left(n_\mathrm{H}+n_\mathrm{He}\right)$ is the free-electron fraction, with $n_e$, $n_\mathrm{H}$ and $n_\mathrm{He}$ the free electron, hydrogen nuclei and helium nuclei number densities, respectively. The first line in Eq.~\eqref{eq: 4} represents the change in temperature in the absence of external heating sources; the first term accounts for adiabatic cooling due to the expansion of the Universe, the second term accounts for adiabatic heating due to clustering of matter and the third term accounts for adiabatic cooling/heating as a result of increasing/lowering the total number of baryons in reionization/recombination. The second line in Eq.~\eqref{eq: 4} represents heating by external sources, with the first (second) term accounting for heating by CMB (X-ray) photons. The expression for Compton heating by CMB photons is well-known and can be found in many papers in the literature, including Refs.~\cite{Peebles:1994xt, Ma:1995ey, Seager:1999bc, Seager:1999km, Barkana:2005xu, Naoz:2005pd}. In contrast, the calculation of X-ray heating, relevant only during the cosmic dawn and reionization epochs, is much more complicated and model dependent, see e.g.~Ref.~\cite{Mesinger:2010ne}.

Another differential equation that has to be solved simultaneously with Eq.~\eqref{eq: 4} is the Boltzmann equation for the free electron fraction $x_e$. The simpler form of this equation can be found in Refs.~\cite{Flitter:2023mjj,Flitter:2023rzv}. {\tt 21cmFirstCLASS} involves the state of the art code {\tt HYREC}~\cite{Ali-Haimoud:2010hou, Lee:2020obi} that solves with great precision the fully-detailed Boltzmann equation for $x_e$ while accounting for helium recombination and high-order energy excitations.

\subsection{Growth factor}\label{subsec: Growth factor}

According to the standard cosmological model, two kinds of matter controlled the evolution of the Universe at early times, before the first atoms were formed. The first kind are baryons, which include hydrogen and helium nuclei and electrons\footnote{We follow standard nomenclature and cavalierly group electrons with baryons.} and protons that interact strongly with photons due to their electric charge. The second kind is collisionless CDM. While the baryons were subject to radiation pressure on small scales, CDM was able to freely cluster under the influence of self-gravitational forces. As the Universe expanded, it eventually reached a low enough temperature such that neutral atoms formed for the first time. This promptly led to a dramatic increase in the photon mean free path, thus releasing the freely propagating CMB radiation. This key moment in the history of the Universe, dated at redshift $z\approx1100$, is known as ``recombination".

\begin{figure*}
\includegraphics[width=\textwidth]{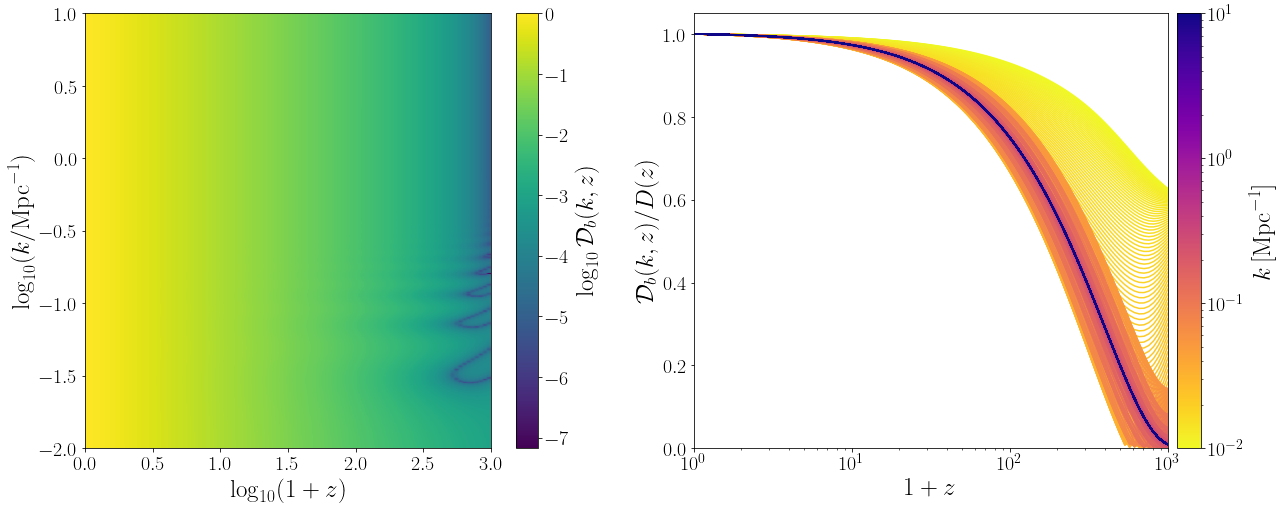}
\caption{Left panel: the scale-dependent growth factor for baryons; note the fluctuations at very high $z$, where the baryon fluid oscillates being still partially coupled to photons. Right panel: comparison with the scale-independent growth factor.}
\label{fig: 1}
\end{figure*}

During the dark ages, the fluctuations in the matter density field were still much smaller than unity and linear perturbation theory can be applied in that epoch. After recombination took place, the linearized continuity and Euler equations for the baryon and CDM fluids on sub-horizon scales are~\cite{baumann_2022, Dodelson:2003ft},
\begin{eqnarray}
\dot\delta_b+a^{-1}\boldsymbol\nabla\cdot\mathbf v_b&=&0\label{eq: 5}
\\\dot{\mathbf v}_b+H\mathbf v_b+a^{-1}\boldsymbol\nabla\Phi&=&-a^{-1}c_s^2\boldsymbol\nabla\delta_b\label{eq: 6}
\\\dot\delta_c+a^{-1}\boldsymbol\nabla\cdot\mathbf v_c&=&0\label{eq: 7}
\\\dot{\mathbf v}_c+H\mathbf v_c+a^{-1}\boldsymbol\nabla\Phi&=&0,\label{eq: 8}
\end{eqnarray}
where $\delta_c$ is the CDM energy over-density, $\mathbf v_c$ is the proper CDM peculiar velocity, and $c_s$ the baryonic sound speed. Overdots denote derivatives with respect to the cosmological time $t$. Eqs.~\eqref{eq: 5}-\eqref{eq: 8} are accompanied with the Poisson equation for the gravitational potential $\Phi$, 
\begin{equation}\label{eq: 9}
\nabla^2\Phi=4\pi G\bar\rho_m a^2\delta_m,
\end{equation}
where $G$ is Newton's gravitational constant, $\bar\rho_m$ is the mean matter energy density and $\delta_m$ the matter energy over-density,
\begin{equation}\label{eq: 10}
\delta_m=\Omega_m^{-1}\left(\Omega_c\delta_c+\Omega_b\delta_b\right).
\end{equation}

Below the comoving Jeans scale, $\sim10\,\mathrm{kpc}$, the fluctuations of the 21-cm signal are suppressed due to smoothing caused by the pressure of the baryonic gas~\cite{Loeb:2003ya, Naoz:2005pd}, as is manifest through the $c_s$ term in Eq.~\eqref{eq: 6}. We have showed in a previous work~\cite{Flitter:2023rzv} that above the Jeans scale, at a wavenumber $k\lesssim50\,\mathrm{Mpc}^{-1}$, the fluctuations in the 21-cm signal can be very well captured by omitting this term. Under this approximation, the equations of motion for the baryon and CDM fluids become identical and we can combine Eqs.~\eqref{eq: 5}-\eqref{eq: 8} into two coupled second-order differential equations of the same form ($i=b,c$ for baryons and CDM),
\begin{equation}\label{eq: 11}
\ddot\delta_i +2H\dot\delta_i-4\pi G\bar{\rho}_{m}\sum_j\frac{\Omega_j}{\Omega_m}\delta_j=0.
\end{equation}
Even though the equations of motion for $\delta_b$ and $\delta_c$ are the same (at the relevant scales), their solutions are different because of the initial conditions at the time of recombination; prior that moment, CDM has already clustered on all scales, while the energy density of the baryons was oscillating 
due to the strong coupling with CMB photons. Thus, $\delta_b\leq\delta_c$ on all scales.

If $\delta_b=\delta_c=\delta_m$, then Eq.~\eqref{eq: 11} becomes a single second order differential equation for $\delta_m$. Since this differential equation is the same on all scales, variable separation can be performed, $\delta_m\left(\mathbf x,z\right)=\delta_0\left(\mathbf x\right)D\left(z\right)$, where $\delta_0\left(\mathbf x\right)$ is a function of only the comoving coordinate $\mathbf x$ in real space, and $D\left(z\right)$ is the scale-independent growth factor (SIGF), a 
function of redshift only that 
obeys the same differential equation for $\delta_m$,
\begin{equation}\label{eq: 12}
\ddot D +2H\dot D-4\pi G\bar{\rho}_{m}D=0,
\end{equation}
with the boundary conditions $D\left(z=0\right)=1$ and $D\left(z\to\infty\right)=0$. Because $\delta_c\geq\delta_b$ and $\Omega_c>\Omega_b$, it is often approximated $\delta_m\approx\delta_c$ and $\delta_c\left(\mathbf k,z\right)=\delta_0\left(\mathbf k\right)D\left(z\right)$ in Fourier space. In this work we use a more precise expression for the baryon and CDM over-densities,\footnote{Here we assume that $\delta_b\approx\delta_c=\delta_0$ at $z=0$, which is very well justified as at this redshift the baryons have already clustered to the gravitational potential wells sourced by CDM.}
\begin{equation}\label{eq: 13}
\delta_i\left(\mathbf k,z\right)=\delta_0\left(\mathbf k\right)\mathcal D_i\left(k,z\right)\qquad i=b,c.
\end{equation}
where $\mathcal D_i\left(k,z\right)$ is the scale-dependent growth factor (SDGF) for species $i$. By construction, it is defined as
\begin{equation}\label{eq: 14}
\mathcal D_i\left(k,z\right)\equiv\frac{\mathcal T_i\left(k,z\right)}{\mathcal T_i\left(k,z=0\right)}\qquad i=b,c.
\end{equation}
where $\mathcal T_i\left(k,z\right)=\delta_i\left(k,z\right)/\mathcal R\left(k\right)$ is the matter density transfer function of species $i$ and $\mathcal R$ is the primordial curvature perturbation.

\begin{figure*}
\includegraphics[width=0.8\textwidth]{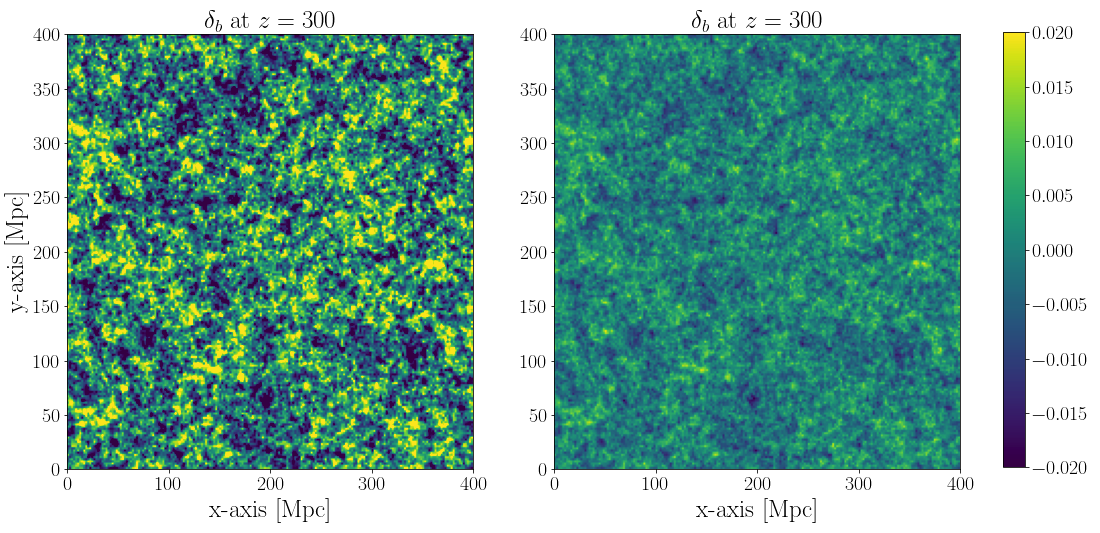}
\caption{Left (right) panel: realization of the CDM (baryon) over-density field at $z=300$. The color bar is the same for the two plots, hence highlighting the stronger fluctuations in the CDM field.}
\label{fig: 2}
\end{figure*}

The CDM SDGF is nearly scale-invariant and can be very well approximated by $\mathcal D_c\left(k,z\right)\approx D\left(z\right)$, apart for BAO features that appear at $z\lesssim1$ (while adopting the SIGF for the CDM density field causes ``fake BAOs" at high redshifts, see more details in Appendix \ref{sec: Scale-dependent matter evolution}). In contrast, the baryon SDGF $\mathcal D_b\left(k,z\right)$ is much smaller than $D\left(z\right)$ at high redshifts, while at low redshifts they are nearly the same due to the increasing  coupling due to gravitation between the baryon and CDM fluids. This is demonstrated in Fig.~\ref{fig: 1}, where BAO features are clearly manifest. Since $\mathcal D_b\left(k,z\right)\ll\mathcal D_c\left(k,z\right)$ at the dark ages, we have $\delta_b\left(\mathbf x,z\right)\ll\delta_c\left(\mathbf x,z\right)$ from Eq.~\eqref{eq: 13}, as illustrated in Fig~\ref{fig: 2}. This has some important implications on the 21-cm signal at that epoch, as we show in Sec~\ref{subsec: Dark ages}.

\subsection{Non linear evolution}\label{subsec: Non linear evolution}

Once galaxies have started to form inside halos, the baryon density field is far from homogeneous as $\delta_b>1$ in regions of galaxy clusters. Thus, during cosmic dawn and reionization epochs linear perturbation theory is no longer applicable in determining $\delta_b\left(\mathbf x,z\right)$. Non-linear perturbation theories, such as standard perturbation theory (or effective field theory)~\cite{Baumann:2010tm, Carrasco:2012cv, Carrasco:2013mua, Hertzberg:2012qn, Porto:2013qua, Senatore:2014eva, Senatore:2014vja, Baldauf:2015aha}, renormalized perturbation theory~\cite{Crocce:2005xy, Bernardeau:2008fa, Montesano:2010qc}, the renormalization group approach~\cite{McDonald:2006hf}, closure theory~\cite{Taruya:2007xy}, TimeRG theory~\cite{Matarrese:2007wc, Pietroni:2008jx}, or time-sliced perturbation theory~\cite{Blas:2015qsi}, must be used in order to find the ``non-linear" perturbation in the baryon density field $\delta_b^\mathrm{NL}$. In this work, we consider a 2nd-order Lagrangian perturbation theory~\cite{1978ApJ...221..937F, Buchert:1989xx, Moutarde:1991evx, 1992MNRAS.254..729B, 1993A&A...267L..51B, Bouchet:1994xp, Hivon:1994qb, Buchert:1995bq, Tatekawa:2004mq, Matsubara:2008wx, Carlson:2009it, Zheligovsky:2013eca} (2LPT).

We provide in Appendix \ref{sec: Scale-dependent Lagrangian perturbation theory} the full derivation of the scale-dependent 2LPT equations. In brief, in this method parcels move from their initial Lagrangian coordinates $\mathbf q$ at redshift $z_0$ to their Eulerian coordinates $\mathbf x$ according the displacement vector field $\mathbf s$, such that $\mathbf x\left(\mathbf q,z\right)=\mathbf q\left(z_0\right)+\mathbf s\left(\mathbf q,z\right)$. The solution to $\mathbf s$ can then be expanded in perturbation theory, $\mathbf s=\mathbf s^{(1)}+\mathbf s^{(2)}$. Under the SIGF approximation, in which $\delta_b=\delta_c$, $\mathbf s^{(1)}$ and $\mathbf s^{(2)}$ are given by
\begin{eqnarray}
\mathbf{s}^{\left(1\right)}\left(\mathbf{q},z\right)&=&\left[D\left(z\right)-D\left(z_{0}\right)\right]\boldsymbol\nabla_{\mathbf{q}}\psi^{\left(1\right)}\left(\mathbf{q}\right)\label{eq: 15}
\\\mathbf{s}^{\left(2\right)}\left(\mathbf{q},z\right)&=&\left[E\left(z\right)-E\left(z_{0}\right)\right]\boldsymbol\nabla_{\mathbf{q}}\psi^{\left(2\right)}\left(\mathbf{q}\right),\label{eq: 16}
\end{eqnarray}
where
\begin{equation}\label{eq: 17}
E\left(z\right)=-\frac{3}{7}D^2\left(z\right),\qquad \nabla_{\mathbf q}^2\psi^{(1)}\left(\mathbf q\right)=-\delta_0\left(\mathbf q\right),
\end{equation}
\begin{equation}\label{eq: 18}
\nabla_{\mathbf{q}}^{2}\psi^{\left(2\right)}=\frac{1}{2}\left[\left(\nabla_{\mathbf{q}}^{2}\psi^{\left(1\right)}\right)^{2}-\left(\frac{\partial^{2}\psi^{\left(1\right)}}{\partial q_{n}\partial q_{m}}\right)^{2}\right].
\end{equation}
In Sec.~\ref{sec: Methods}, we discuss how these equations change when the SDGF is introduced.

\subsection{Conditional mass function}\label{subsec: Conditional mass function}

During the cosmic dawn and reionization epochs, the 21-cm signal does not depend only on the baryon over-density $\delta_b$, but also on the CDM over-density field $\delta_c$, or more precisely on the total matter over-density $\delta_m$, Eq.~\eqref{eq: 10}. This is because galaxies and stars, which emit crucial radiation that alters the signal, form inside collapsed dark matter halos. In analytical and semi-numerical models, the effect of the halos on the signal comes into play through the conditional halo mass function (HMF),  $dn\left(\mathbf x,M_h,z\right)/dM_h$; it is conditional in the sense that it depends on the value of the local total matter over-density $\delta_m\left(\mathbf x,z\right)$. The mean of the conditional HMF is the ``standard" HMF, defined as~\cite{Lukic:2007fc}
\begin{equation}\label{eq: 19}
\frac{d\bar n_h\left(M_h,z\right)}{dM_h}=\frac{\bar\rho_{m,0}}{M_h}\frac{d\ln\sigma^{-1}\left(M_h,z\right)}{dM_h}f\left(\sigma\right),
\end{equation}
where $\bar\rho_{m,0}$ is the background value of the current (at $z=0$) total matter density field and $\sigma^2\left(M_h,z\right)$ is the variance of the \emph{linear} total matter density field, smoothed with a top-hat filter in real space of comoving radius $R_{M_h}=\left(3M_h/4\pi\bar\rho_{m,0}\right)^{1/3}$,
\begin{equation}\label{eq: 20}
\sigma^2\left(M_h,z\right)=\int_0^\infty\frac{dk}{k} A_s\left(\frac{k}{k_\star}\right)^{n_s-1}\mathcal T^2_m\left(k,z\right)W^2\left(kR_{M_h}\right),
\end{equation}
where $\mathcal T_m\left(k,z\right)\equiv\delta_m^\mathrm{lin}\left(k,z\right)/\mathcal R\left(k\right)$ is the linear matter density transfer function and $W\left(kR\right)=3\left(\sin kR-kR\cos kR\right)/\left(kR\right)^3$. The function $f\left(\sigma\right)$ is the HMF fitting function and is calibrated on data from N-body simulations. Perhaps the simplest choice is to rely on the purely Gaussian fitting function from the Press-Schechter (PS)~\cite{Press:1973iz} HMF. While the simple form of the PS fitting function captures well the qualitative behavior of the HMF, it under-estimates the number of massive halos. For that reason, we use in this work instead the Sheth-Tormen (ST) HMF~\cite{Sheth:1999mn, Sheth:1999su, Sheth:2001dp} (but see other popular mass functions in Refs.~\cite{Jenkins:2000bv, Warren:2005ey, Reed:2006rw, Tinker:2008ff, Watson:2012mt}). 

The calculation of the emissivity of the radiating sources often involves the conditional halo mass function through the integral in Eq.~\eqref{eq: 25}. In evaluating that integral, the following quantity can be useful,
\begin{equation}\label{eq: 21}
f_\mathrm{coll}\left(\mathbf x,z;M_\mathrm{min}\right)=\frac{1}{\bar\rho_{m,0}}\int_{M_\mathrm{min}}^\infty dM_h\,M_h\frac{dn_h\left(\mathbf x,z,M_h\right)}{dM_h}.
\end{equation}
This is the \emph{local} collapsed fraction of halos more massive than $M_\mathrm{min}\left(z\right)$. From Eq.~\eqref{eq: 21} it is clear then that the conditional HMF can be extracted from the collapsed fraction through the relation 
\begin{equation}\label{eq: 22}
\frac{dn\left(\mathbf x,z,M_h\right)}{d M_h}=-\frac{\bar{\rho}_{m,0}}{M_h}\frac{df_{\mathrm{coll}}\left(\mathbf x,z;M_h\right)}{dM_h}.
\end{equation}
The collapsed fraction is a very useful quantity since the extended Press-Schechter (EPS) theory predicts its value analytically~\cite{Bond:1990iw}. According to EPS, for a PS HMF, in a large-scale region of comoving radius $R$, centered at $\mathbf x$, with a smoothed over-density $\delta_{m,R}$, the collapsed fraction of halos above $M_\mathrm{min}$ at redshift $z$ is
\begin{flalign}\label{eq: 23}
\nonumber &f_\mathrm{coll}^\mathrm{EPS}\left(\mathbf x,z;M_\mathrm{min}\right)=&
\\&\hspace{25mm}\mathrm{erfc}\left(\frac{\delta_\mathrm{crit}-\delta_{m,R}\left(\mathbf x,z\right)}{\sqrt{2\left[\sigma^2\left(M_\mathrm{min},z\right)-\sigma^2\left(M_R,z\right)\right]}}\right),&
\end{flalign}
where $\mathrm{erfc}\left(\cdot\right)$ is the complementary error function, $\delta_\mathrm{crit}=1.686$, and 
\begin{eqnarray}\label{eq: 24}
\nonumber M_R&=&\frac{4\pi}{3}R^3\bar\rho_{m,0}
\\&=&1.66\times10^{11}\left(\frac{h^{2}\Omega_{m}}{0.143}\right)\left(\frac{R}{1\,\mathrm{Mpc}}\right)^{3}\,M_{\odot}.
\end{eqnarray}
Note that in writing Eq.~\eqref{eq: 23} we have used the redshift-dependent $\sigma\left(M_h,z\right)$ as given by Eq.~\eqref{eq: 20}, without imposing the SIGF approximation, $\sigma\left(M_h,z\right)\propto D\left(z\right)$, which is less accurate at high redshifts and low masses (see more details in Appendix \ref{sec: Scale-dependent matter evolution}).

One delicate nuance regarding the formalism presented above concerns the assumed matter density field that is used in Eq.~\eqref{eq: 23}. Press-Schechter theory~\cite{Sheth:1999mn}, and the extended Press-Schechter theory~\cite{Bond:1990iw}, are both established on \emph{linear} cosmological perturbation theory. Hence, from a purely theoretical point of view, it would be more consistent to evaluate the smoothed $\delta_m$ that appears in Eq.~\eqref{eq: 23}, as well as $\sigma\left(M_h,z\right)$, from linear theory. This is for example the underlying assumption in {\tt Zeus21}, where the ``linear" $\delta_m^\mathrm{lin}$ still results in non-linear fluctuations in the SFRD due to the erfc function in Eq.~\eqref{eq: 23}, an effect that can be captured by an exponential fit~\cite{Munoz:2023kkg}. In contrast, {\tt 21cmFAST} uses the non-linear $\delta_m^\mathrm{NL}$ (from 2LPT calculations) in Eq.~\eqref{eq: 23}. The justification for this choice comes from better agreement with radiative transfer simulations at the epoch of reionization~\cite{Zahn:2010yw}.\footnote{We comment that the non-linear density field contains non-Gaussian statistics and therefore non-vanishing bispectrum, leading to a much more complicated $f_\mathrm{coll}$, see Eq.~(5) in Ref.~\cite{Lidz:2013tra}.}

\subsection{Star formation rate density}\label{subsec: Star formation rate density}

The role of the baryons in shaping the 21-cm signal is enhanced when the dark ages are over. As dark matter halos collapse, the gravitational pull that the baryons experience is increased. If the cooling timescale is shorter than the free-fall timescale, zero-metalicity popIII stars can form via gas accretion, mostly in molecular cooling galaxies (MCGs, other works often refer to molecular cooling halos for more generality~\cite{Tegmark:1996yt, Abel:2001pr, Bromm:2003vv, Haiman:2006si, Trenti:2010hs, Plombat:2024exw}). Baryonic feedbacks, such as supernovae~\cite{Springel:2002ux, Wise:2007nb, Pawlik:2008dk}, photo-heating~\cite{Hui:1997dp, Okamoto:2008sn, Ocvirk:2018pqh, Katz:2019due} and Lyman-Werner radiation~\cite{Bromm:2003vv, Fialkov:2012su, Haiman:1996rc} hamper the star formation rate (SFR). Still, stars continue to form in more massive halos that host atomic cooling galaxies (ACGs)~\cite{Qin:2020xyh, Qin:2020pdx, Krumholz:2012wi}. Those halos typically are assumed to contain mainly popII stars with low metallicity, though popIII stars can still form in atomic cooling halos if there is sterilizing radiation background that prevents star formation earlier in the molecular cooling stage (when the halo was smaller and younger).

The impact that stars have on the 21-cm signal is immense. First, Ly$\alpha$ radiation couples the spin temperature to the gas kinetic temperature via the Wouthuysen-Field (WF) effect~\cite{1952AJ.....57R..31W, 1958PIRE...46..240F} (this ``turns on" $\tilde x_\alpha$ in Eq.~\ref{eq: 3}). Secondly, X-ray radiation heats up the gas, increasing the value of $T_k$ (see last term in Eq.~\ref{eq: 4}). And thirdly, UV photons ionize the IGM, thus lowering the value of $\tau_{21}$ significantly. Those three effects are controlled by the fluxes $J_\alpha$, $J_X$ and $J_\mathrm{UV}$. These radiation
fluxes depend on non-local
values of the emissivity field, which can be related to the star formation rate density (SFRD) field under certain assumptions. The inhomogeneous SFRD field is then closely related with the baryon density field, as galaxies are biased tracers of the inhomogeneous baryon density field, as can be seen by its mathematical expression~\cite{Madau:1996cs, Schneider:2020xmf, Munoz:2021psm, Munoz:2023kkg, Cruz:2024fsv},
\begin{equation}\label{eq: 25}
\dot\rho_*\left(\mathbf x,z\right)=\int dM_{h}\frac{dn_{h}\left(\mathbf{x},M_{h},z\right)}{dM_{h}}\dot{M}_{*}\left(\mathbf{x},M_{h},z\right),
\end{equation}
where $\dot{M}_{*}\left(\mathbf{x},M_{h},z\right)$ is the local stellar mass production rate in a halo of mass $M_h$. Clearly, the local SFRD is smaller in regions with smaller $\dot{M}_{*}$, while the latter quantity must be considerably low at halos of mass $\sim M_\mathrm{min}$, where the halo mass is insufficient for efficient cooling. In practice, Eqs.~\eqref{eq: 22} and \eqref{eq: 23} are used in computing $\dot\rho_*\left(\mathbf x,z\right)$ under the assumption of PS HMF, while the unconditional ST HMF is then used to normalize the mean SFRD~\cite{Barkana:2003qk, Barkana:2007xj, Mesinger:2010ne}.

The local SFRD does not depend only on the local baryon density, but also on local velocities, specifically on the relative velocity between baryons and CDM, $\mathbf v_{cb}=\mathbf v_c-\mathbf v_b$. From Eqs.~\eqref{eq: 6} and \eqref{eq: 8}, we see that above the Jeans scale $\dot{\mathbf{v}}_{cb}=-H\mathbf{v}_{cb}$, which has the temporal solution $\mathbf{v}_{cb}\left(\mathbf x,z\right)\propto\left(1+z\right)$. As for the spatial solution, adiabatic initial conditions imply Gaussian distributions for each one of the three components of $\mathbf{v}_{cb}$, and a Maxwell-Boltzmann distribution for $v_{cb}\equiv\left(\mathbf v_{cb}\cdot\mathbf v_{cb}\right)^{1/2}$. The variance of this distribution can be determined by defining $\theta_{cb}\equiv a^{-1}\boldsymbol\nabla\cdot\mathbf v_{cb}$, and then
\begin{equation}\label{eq: 26}
\langle v_{cb}^2\left(z\right)\rangle=\int_0^\infty\frac{dk}{k} A_s\left(\frac{k}{k_\star}\right)^{n_s-1}\frac{\mathcal T^2_{\theta_{cb}}\left(k,z\right)}{k^2},
\end{equation}
where $\mathcal T_\mathrm{\theta_{cb}}\left(k,z\right)=\theta_{cb}\left(k,z\right)/\mathcal R\left(k\right)$ is the transfer function for $\theta_{cb}$. Evaluating this integral at recombination ($z\approx1100$) reveals that baryons move with a mean relative velocity of $\sim 30\,\mathrm{km/sec}$ with respect to CDM, corresponding to a Mach number of 5~\cite{Tseliakhovich:2010bj, Ali-Haimoud:2013hpa}. This supersonic velocity makes it harder for the baryons to collapse and form popIII stars in small dark matter halos~\cite{Tseliakhovich:2010yw, Dalal:2010yt, Fialkov:2011iw, Greif:2011iv, Stacy:2010gg, OLeary:2012gem, Schauer:2018iig, Hirano:2017znw, Barkana:2016nyr, Munoz:2019rhi}, thereby lowering the SFRD and delaying cosmic dawn. 

While the quantitive effect of $v_{cb}$ on the SFRD is model-dependent, the qualitative impact on the 21-cm signal is not. Less stars (with fixed luminosity) imply weaker X-ray and Ly$\alpha$ fluxes, resulting in a delay in the 21-cm signal. In {\tt 21cmFAST} and {\tt 21cmFirstCLASS} this effect is modelled in the code by increasing $M_\mathrm{min}$ in MCGs~\cite{Munoz:2021psm},
\begin{equation}\label{eq: 27}
M_\mathrm{min}^\mathrm{MCG}\propto\left[1+A_{v_{cb}}\frac{v_{cb}\left(\mathbf x,z_\mathrm{rec}\right)}{\langle v_{cb}^2\left(z_\mathrm{rec}\right)\rangle^{1/2}}\right]^{\beta_{v_{cb}}},
\end{equation}
where $A_{v_{cb}}$ and $\beta_{v_{cb}}$ are free parameters, and $\langle v_{cb}^2\left(z_\mathrm{rec}\right)\rangle$ is the variance of the $v_{cb}$ field at recombination (see Eq.~\ref{eq: 26}). This modification in $M_\mathrm{min}$ results in having VAOs in the 21-cm power spectrum at large scales~\cite{Munoz:2019rhi, Munoz:2019fkt, Sarkar:2022mdz, Zhang:2024pwv}, as was first studied in Refs.~\cite{Dalal:2010yt, Visbal:2012aw, Fialkov:2012su}.

\section{Methods}\label{sec: Methods}

The main tool we use in this work to study how the 21-cm signal is sensitive to the assumed underlying matter fields is {\tt 21cmFirstCLASS}. This code interfaces {\tt CLASS}~\cite{Blas:2011rf} and {\tt 21cmFAST}, allowing the simulation to begin from recombination with initial conditions that are consistent with the simulated cosmology, evolve the matter fields throughout the dark ages, and apply 2LPT during cosmic dawn and reionization to simulate the non-linear fluctuations of the signal in those epochs. Moreover, the merger between {\tt CLASS} and {\tt 21cmFAST} enables a consistent study of both CMB and 21-cm power spectra with {\tt 21cmFirstCLASS}, even for cosmologies that extend beyond $\Lambda$CDM~\cite{Flitter:2023mjj, Adi:2024ebl, Libanore:2025ack}.

For this work, three major modifications have been made in {\tt 21cmFirstCLASS} with respect to its previous version, whose full description can be found in Ref.~\cite{Flitter:2023mjj}. Firstly, we enabled the option of using the linear matter density field $\delta_m^\mathrm{lin}$, rather the non-linear matter density field $\delta_m^\mathrm{NL}$ in the evaluation of the inhomogeneous SFRD $\dot\rho_*\left(\mathbf x,z\right)$ (see discussion at the end of Sec.~\ref{subsec: Conditional mass function}). Secondly, we changed the effect that $v_{cb}$ has on the local SFRD; in addition to the modification of $M_\mathrm{min}^\mathrm{MCG}$ by $v_{cb}$ (Eq.~\ref{eq: 27}), {\tt 21cmFAST} also modifies the current matter power spectrum by introducing a multiplicative correction factor of the form $1-A_p\exp\left[-\left(\ln\left[k/k_p\right]\right)^2/\left(2\sigma_p^2\right)\right]$~\cite{Munoz:2021psm}. This factor arises from non-linear $v_\mathrm{cb}$ corrections, and was calibrated at $z=20$ with $A_p=0.24$, $k_p=300\,\mathrm{Mpc}^{-1}$, and $\sigma_p=0.9$. {\tt 21cmFirstCLASS} has the option to remove this factor since it introduces non-linear corrections to $\delta_m^\mathrm{lin}$ and $\sigma\left(M_h,z\right)$, while the EPS formalism requires these quantities to be evaluated from linear theory, as we now do in {\tt 21cmFirstCLASS}.

\begin{figure}
\includegraphics[width=\columnwidth]{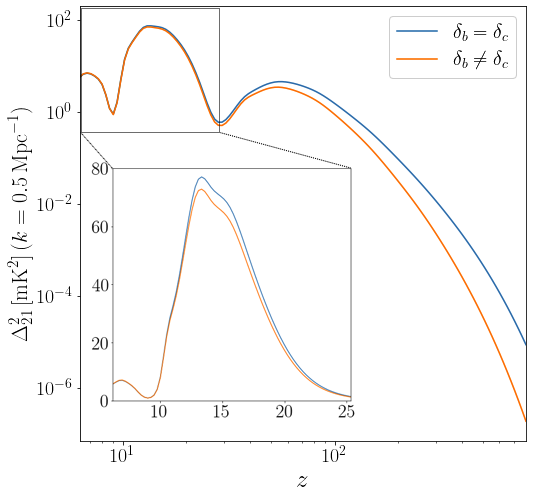}
\caption{The 21-cm power spectrum as a function of redshift at $k=0.5\,\mathrm{Mpc}^{-1}$, when the baryon over-density field is evolved with the SIGF (blue curve) or the SDGF (orange curve). The inset shows a zoomed-in version of the curves after the dark ages, where a maximum difference of 6\% can be seen.}
\label{fig: 3}
\end{figure}

\begin{figure*}
\includegraphics[width=0.8\textwidth]{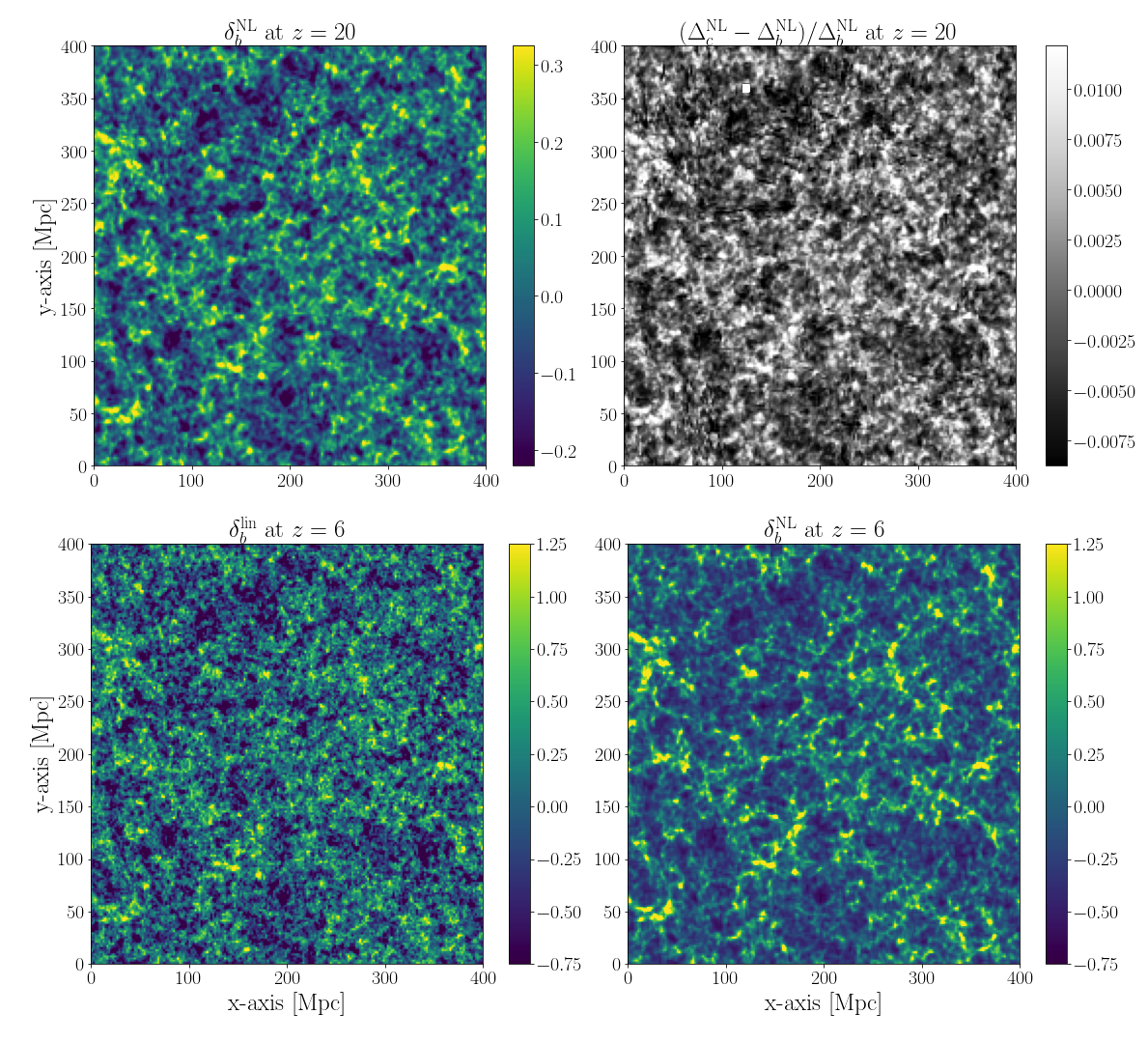}
\caption{\textbf{Top left panel}: realization of the baryon over-density field at $z=20$, evolved non-linearly with scale-dependent 2LPT. \textbf{Top right panel}: the fractional error in the normalized density field ($\Delta_i\equiv1+\delta_i$) due to approximating $\delta_b\approx\delta_c$. The error is of the order of a percent at the most over- and under-dense regions. \textbf{Bottom left (right) panel}: realization of the baryon over-density field at $z=6$ when evolved linearly (non-linearly). Linear perturbation theory fails to capture the clustering of baryons at low redshifts and small scales.}
\label{fig: 4}
\end{figure*}

The third modification that was made in {\tt 21cmFirstCLASS} is relaxing the SIGF approximation in the calculations by accounting for the more exact SDGF. Assuming that linear perturbation theory holds during the dark ages, we have shown in Ref.~\cite{Flitter:2023rzv} that implementing the SDGF in {\tt 21cmFirstCLASS} would yield excellent agreement with fully relativistic Boltzmann-Einstein solvers such as {\tt CAMB}, on sub-horizon scales that are larger than the Jeans scale (we verify the correctness of our calculations in Appendix \ref{sec: Temperature fluctuations in the dark ages}). On top of this modification during the dark ages, we improved the 2LPT calculations in {\tt 21cmFAST} to account for scale-dependent growth also in the non-linear epochs of cosmic dawn and reionization. When relaxing the SIGF approximation, the displacement vector field depends on the considered matter species, and we denote it by $\mathbf s_i\left(\mathbf q,z\right)$ ($i=b,c$). Now, the first order correction $\mathbf s_i^{(1)}\left(\mathbf q,z\right)$ cannot be decomposed to spatial and temporal components, as in Eq.~\eqref{eq: 15}, and instead we must compute it in Fourier space,
\begin{equation}\label{eq: 28}
\mathbf{s}_{i}^{\left(1\right)}\left(\mathbf{k},z\right)=\left[\mathcal{D}_{i}\left(k,z\right)-\mathcal{D}_{i}\left(k,z_{0}\right)\right]\frac{i\mathbf{k}}{k^{2}}\delta_{0}\left(\mathbf{k}\right).
\end{equation}
When we relax the SIGF approximation in {\tt 21cmFirstCLASS}, we use Eq.~\eqref{eq: 28} for $\mathbf s_i^{(1)}$, instead of Eq.~\eqref{eq: 15}. On the other hand, the second order correction, $\mathbf s_i^{(2)}$, should be found by solving a set of coupled integro-differential equations, that the interested reader can find in Appendix \ref{sec: Scale-dependent Lagrangian perturbation theory}, Eq.~\eqref{eq: C22}. Instead, in {\tt 21cmFirstCLASS} we keep using the approximation in Eq.~\eqref{eq: 16}; as described in detail in Appendix \ref{sec: Scale-dependent Lagrangian perturbation theory}, this is justified as by construction the second order correction $\mathbf s_i^{(2)}$ has a smaller contribution to $\mathbf s_i$, compared to the first order correction $\mathbf s_i^{(1)}$. We leave for future work the comparison of the approximated analytical solution of Eq.~\eqref{eq: 16} with the fully numerical solution of Eq.~\eqref{eq: C22}.

\section{Results}\label{sec: Results}

\subsection{Dark ages}\label{subsec: Dark ages}

During the dark ages, the only relevant matter species that contributes to the 21-cm signal are the baryons, as they source the emission and absorption of 21-cm radiation by hydrogen atoms, the recombination rate that governs $T_k$, and the coupling of $T_s$ to $T_k$ via collisional excitations. Thus the fluctuations in the 21-cm signal during the dark ages are entirely proportional to $\delta_b$. Because $\delta_b\ll1$ during the dark ages, the scale-dependent evolution of the baryon density field barely affects the global 21-cm signal, i.e. the sky-averaged brightness temperature as a function of redshift, $\bar T_{21}\left(z\right)$. Indeed, for the cosmological and astrophysical parameters assumed in this work, we find that the SIGF approximation yields a maximum error of 0.4\% in $\bar T_{21}$ at $z\simeq14$, where the global signal reaches its minimum value.

Although the scale-dependent evolution of $\delta_b$ does not affect the global 21-cm signal, it does affect its two-point statistics. Specifically, let us consider the 21-cm power spectrum,
\begin{equation}\label{eq: 29}
\Delta_{21}^2\left(k,z\right)=\frac{k^3\bar T_{21}^2\left(z\right)P_{21}\left(k,z\right)}{2\pi^2},
\end{equation}
where $P_{21}\left(k, z\right)$ is the angle-averaged Fourier transform of the two-point correlation function $\langle\delta_{21}\left(\mathbf x, z\right) \delta_{21}\left(\mathbf x', z\right)\rangle$, while $\delta_{21}$ is the local contrast in the brightness temperature, $\delta_{21}\left(\mathbf x, z\right)\equiv T_{21}\left(\mathbf x, z\right)/\bar T_{21}\left(z\right)-1$. In a previous work~\cite{Flitter:2023rzv}, we have shown that during the dark ages $\Delta_{21}^2\left(k,z\right)\propto\mathcal T_b^2\left(k,z\right)\ll\mathcal T_c^2\left(k,z\right)$.
In Fig.~\ref{fig: 3}, we show $\Delta_{21}^2\left(k,z\right)$, as calculated by {\tt 21cmFirstCLASS} after the incorporation of the SDGF in the code. As expected, $\Delta_{21}^2\left(k,z\right)$ is suppressed during the dark ages when $\delta_b$ is correctly evolved with the SDGF. Specifically, we find that the SIGF approximation yields a $\sim40\%$ error in the 21-cm power spectrum at $z=100$ for $k\gtrsim0.1\,\mathrm{Mpc}^{-1}$.

\subsection{Cosmic dawn and reionization}\label{subsec: Cosmic dawn and reionization}

As was previously discussed in Sec.~\ref{sec: Theory}, when the Universe enters the cosmic dawn epoch, the dependence of the 21-cm signal on the matter fields becomes more complicated as (1) non-linear fluctuations in $\delta_b$ become non-negligible and (2) both $\delta_b$ and $\delta_m$ need to be taken into account.

\begin{figure}
\includegraphics[width=\columnwidth]{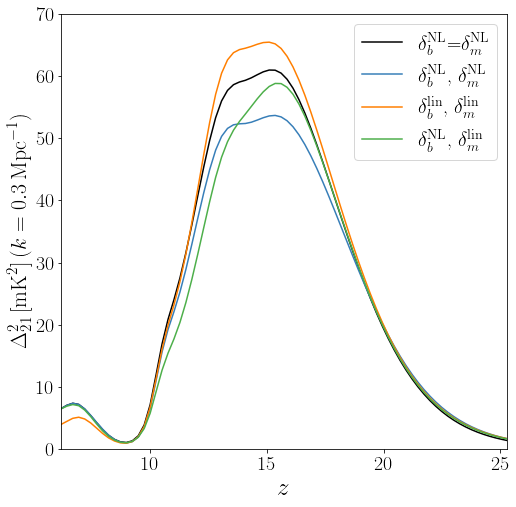}
\caption{The 21-cm power spectrum as a function of redshift at $k=0.3\,\mathrm{Mpc}^{-1}$, when the baryon and the total matter over-density fields are evolved either linearly or non-linearly. The black curve corresponds to adopting the SIGF approximation in which $\delta_b=\delta_c=\delta_m$. The colors of the curves here match the same colors in Figs.~\ref{fig: 6}-\ref{fig: 7}, namely the blue, orange and green curves can be associated with {\tt 21cmFAST}, {\tt Zeus21}-like, and {\tt 21cmFirstCLASS}, respectively. Here though, unlike Figs.~\ref{fig: 6}-\ref{fig: 7}, in all colorful curves we distinguish between baryons and CDM (i.e.~in all colorful curves we use the SDGF), and we have the same $v_{cb}$ delaying effect (see more information at Sec.~\ref{sec: Methods}).}
\label{fig: 5}
\end{figure}

We show in the top panels of Fig.~\ref{fig: 4} the comparison between the more precise $\delta_b^\mathrm{NL}$, computed with the scale-dependent first order LPT, and the common calculation that evolves the density field with the SIGF. At $z=20$, we see that the error is of the order of a percent at the most over- and under-dense regions. This error is expected to increase at higher redshifts due to the departure of the SDGF from the SIGF, while at lower redshifts it vanishes as $\mathcal D_b\left(k,z\right)\to D\left(z\right)$, see Fig.~\ref{fig: 1}. The error in the evaluation of $\delta_b^\mathrm{NL}$ with the SIGF is enhanced when the 21-cm power spectrum is considered, due to the intricate interplay between non-linear fluctuations in different fields, e.g. $T_k$ and $J_\alpha$. Indeed, we see from comparing between the black and blue curves in Fig.~\ref{fig: 5} that the error in $\Delta_{21}^2$ may reach to $\mathcal O\left(10\%\right)$ at $z\gtrsim12$. At low redshifts ($z\lesssim10$), the error due to this approximation goes to zero because at these redshift $\delta_b$ approaches more closely to $\delta_c$.

The importance of evolving the baryon density field non-linearly is evident in the bottom panels of Fig.~\ref{fig: 4}; while linear perturbation theory seems to give a rough estimate for the distribution of baryonic matter on very large scales, it fails to capture the clustering of baryons at low redshifts and small scales, and it wrongly predicts a more homogeneous density field. When computing the 21-cm power spectrum, the imprecise calculation from linear perturbation theory yields an order unity error at $z\lesssim 8$ at $k=0.3\,\mathrm{Mpc}^{-1}$, as demonstrated in Fig.~\ref{fig: 5} (this effect however is also manifest on other scales, as can be seen in Fig.~\ref{fig: 7}). Importantly, this is the regime where upcoming 21-cm interferometers are mostly sensitive (c.f.~Fig.~\ref{fig: 7}).

Finally, Fig.~\ref{fig: 5} also demonstrates how $\Delta_{21}^2\left(k,z\right)$ is modified due to the choice of working with either $\delta_m^\mathrm{lin}$ or $\delta_m^\mathrm{NL}$ in Eq.~\eqref{eq: 23}. Evidently, because the fluctuations in the brightness temperature are strongly connected to the fluctuations in the SFRD,  this choice has large implications on the resulting 21-cm power spectrum during cosmic dawn.

\begin{figure}
\includegraphics[width=\columnwidth]{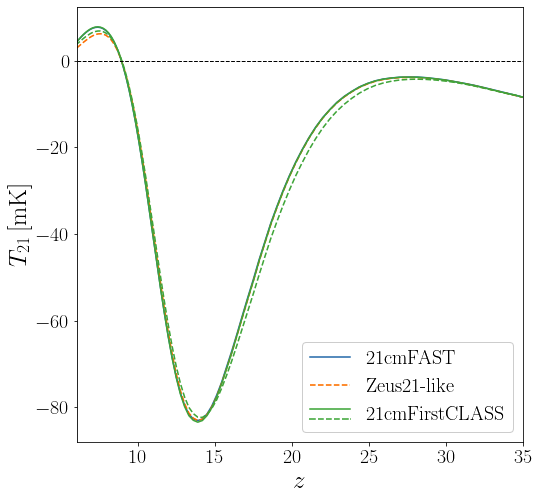}
\caption{The global 21-cm signal under different assumptions. Green lines are obtained using \texttt{21cmFirstCLASS} and separate the baryons and dark matter evolution; other lines assume $\delta_b \equiv \delta_c$. In the blue curve, we ran {\tt 21cmFAST} under the default settings of non-linear evolution; the
orange curve instead is obtained  
setting the density field to evolve linearly, similarly as in {\tt Zeus21} (the two codes well agree in these settings). 
\emph{Solid (dashed)} lines correspond to codes that adopt the non-linear (linear) matter density field in their calculations. In both green curves of {\tt 21cmFirstCLASS}, the baryon density field is evolved non-linearly.}
\vspace{-0.15in}
\label{fig: 6}
\end{figure}

\begin{figure}
\includegraphics[width=0.88\columnwidth]{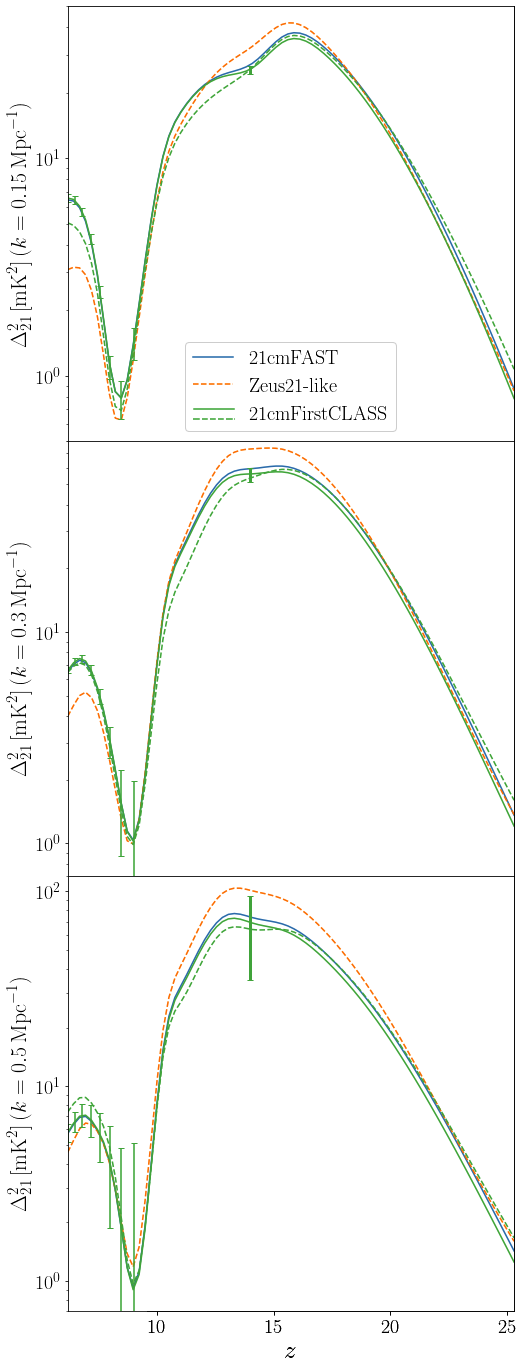}
\caption{Comparison of the 21-cm power spectrum on different scales, under the same conditions as in Fig.~\ref{fig: 6}. Here, we show with error bars the expected noise from HERA in its design sensitivity at low redshifts ($z<10$) assuming the moderate foreground scenario, as evaluated by the {\tt 21cmSense} code~\cite{Pober:2012zz, Pober:2013jna} (we used the same {\tt 21cmSense} settings as detailed in Ref.~\cite{Flitter:2023mjj}). We also show a separate thick error bar at $z=14$, corresponding to a sensitivity estimate for SKA~\cite{Meriot:2023usy}.}
\label{fig: 7}
\end{figure}

\newpage
\section{Discussion}\label{sec: Discussion}

The results shown in the previous section demonstrate that differences in the simulated 21-cm signal can indeed arise by simply using different matter field configurations in the simulation. We now examine if these differences could be significant in future detections during cosmic dawn and reionization epochs. We thus compare between {\tt 21cmFirstCLASS} and two popular codes used in the literature, {\tt 21cmFAST} and {\tt Zeus21}. Similarly to {\tt 21cmFirstCLASS}, {\tt 21cmFAST} performs 2LPT in order to compute the non-linear fluctuations in the matter fields, however there is no distinction between baryons and CDM and the code assumes $\delta_b=\delta_c=\delta_m$. {\tt Zeus21} also does not make the distinction between baryons and CDM, but unlike {\tt 21cmFAST} it assumes the linear matter density field $\delta_m^\mathrm{lin}$ throughout the calculation. A summary of the differences between the codes is given in Table ~\ref{tab: 1}.

\begin{table*}[t!]
\begin{tabular}{|cccccc|}
\hline
Feature & {\tt 21cmFAST} & {\tt Zeus21} & {\tt 21cmFirstCLASS} & Diff. in $\Delta_{21}^2\left(k,z\right)$ & Diff. in $\Delta_{21}^2\left(k,z\right)$ \\
& & & & (cosmic dawn and reionization) & (dark ages)\\
\hline\hline
Non-linear evolution & \ding{52} & \ding{55} & \ding{52} & $\sim20\%\left(z=15,k=0.3\,\mathrm{Mpc}^{-1}\right)$ & --- \\
Scale dependent growth & \ding{55} & \ding{55} & \ding{52} & $\sim10\%\left(z=15, k=0.3\,\mathrm{Mpc}^{-1}\right)$ & $\sim40\%\left(z=100, k=0.3\,\mathrm{Mpc}^{-1}\right)$ \\
Linear theory in EPS & \ding{55} & \ding{52} & \ding{52} & $\sim20\%\left(z=7, k=0.1\,\mathrm{Mpc}^{-1}\right)$ & ---\\
\hline
\end{tabular}
\caption{Summary of features studied in this work, comparing between three public codes: {\tt 21cmFAST}, {\tt Zeus21} and {\tt 21cmFirstCLASS}. The leftmost column shows the three features discussed in this work. \textbf{``Non-linear evolution"} corresponds to perturbing the total matter density field non-linearly, $\delta=\delta^\mathrm{NL}$. \textbf{``Scale dependent growth"} corresponds to the relaxation of the SIGF approximation, $\delta_b\neq\delta_c$. \textbf{``Linear theory in EPS"} corresponds to having $\delta_m=\delta_m^\mathrm{lin}$ in Eq.~\eqref{eq: 23} while removing the non-linear $v_{cb}$ term in the calculation of $\sigma\left(M_h,z\right)$ (see more details in Sec.~\ref{sec: Methods}). The middle columns display whether or not these features are present in the three codes. The two rightmost columns show the relative difference in the 21-cm power spectrum due to switching off the features. During the dark ages, only the second feature affects the 21-cm power spectrum, thus {\tt 21cmFirstCLASS} is the only code among the three that can simulate that epoch. The redshifts of our quoted benchmarks correspond to the minima of the global signal at $z=15$ and $z=100$, or the maximum at $z=7$. Before reionization ($z\gtrsim7$), the relative differences do not vary significantly with scale between $0.1\leq k\leq0.6\,\mathrm{Mpc}^{-1}$.}
\label{tab: 1}
\end{table*}

The comparison between the codes is shown in Figs.~\ref{fig: 6}-\ref{fig: 7}, where {\tt 21cmFAST} ({\tt Zeus21}) is represented with blue (orange) curve. As {\tt 21cmFirstCLASS} is able to use either $\delta_m^\mathrm{NL}$ or $\delta_m^\mathrm{lin}$ (while always using the non-linear $\delta_b^\mathrm{NL}$ in the simulation after the dark ages are over), these cases are represented by the solid and dashed green curves, respectively. We note that in order to compare apples to apples, we obtained the orange ``{\tt Zeus21}-like" curve by running {\tt 21cmFAST} with linear perturbation theory, which was shown to agree well with the {\tt Zeus21} code in Refs.~\cite{Munoz:2023kkg, Cruz:2024fsv}.

In all three codes, the evaluated global signal is roughly the same, though minor differences can still be observed in Fig.~\ref{fig: 6}. When {\tt 21cmFirstCLASS} runs with the $\delta_m^\mathrm{lin}$ configuration, the signal is slightly shifted to higher redshifts, as $\sigma\left(M_h,z\right)$ is calculated directly from linear perturbation theory, Eq.~\eqref{eq: 20}, without the non-linear $v_{cb}$ modification that suppresses power at small scales (see more details in Sec.~\ref{sec: Methods}). Furthermore, the difference in the emission signal during the reionization epoch can get up to 12\% when comparing {\tt 21cmFirstCLASS} with $\delta_m^\mathrm{lin}$ to {\tt 21cmFAST} (or to {\tt 21cmFirstCLASS} with $\delta_m^\mathrm{NL}$). This discrepancy increases to 20\% when {\tt Zeus21} is compared to {\tt 21cmFAST}. These differences arise due to the highly non-linear fluctuations in that epoch; the fluctuations in the signal are so large that they do not cancel out when the mean is considered.

Next, in Fig.~\ref{fig: 7} we compare $\Delta_{21}^2\left(k,z\right)$, as evaluated by the three codes. In this figure, the $z\lesssim10$ error bars indicate the expected noise level from HERA in its design sensitivity under the conservative assumption of a moderate foreground scenario, evaluated with {\tt 21cmSense}~\cite{Pober:2012zz, Pober:2013jna}, while the $z=14$ error bar illustrates the sensitivity expected for SKA, estimated by Ref.~\cite{Meriot:2023usy}. On large scales ($k\lesssim0.3\,\mathrm{Mpc}^{-1}$), the 21-cm power spectrum features the same differences in the global signal at low redshifts, and the type of the assumed matter field ($\delta_m^\mathrm{NL}$ or $\delta_m^\mathrm{lin}$) is crucial for accurately simulating the signal within the noise level of HERA. This conclusion also holds for small scales ($k\gtrsim0.3\,\mathrm{Mpc}^{-1}$), though to a lesser extent. Interestingly, at intermediate scales of $k\sim0.3\,\mathrm{Mpc}^{-1}$ the type of the assumed matter field less matters, suggesting that at these scales the signal mostly depends on $\delta_b$ rather than $\delta_m$, though we comment that this conclusion might change if different astrophysical parameters are considered. In all scales, the usage of $\delta_b^\mathrm{lin}$ is inappropriate for simulating the signal within HERA sensitivity range, especially during the epoch of reionization. At higher redshifts, we see that distinguishing $\delta_b$ from $\delta_c$ causes differences in $\Delta_{21}^2\left(k,z\right)$ on large scales that are comparable to the expected SKA noise level.

Fig.~\ref{fig: 7} thus demonstrates that the subtleties discussed in this work should matter in future analyses of  21-cm signal measurements. Yet, perhaps the most crucial subtlety in simulating the signal may come from another nuance which is treated the same in all three codes discussed here and has to do with the appearance of the smoothing radius $R$ in the RHS of Eq.~\eqref{eq: 23}. According to EPS, this $R$ should correspond to the region in which the local collapsed fraction is evaluated. Hence, we argue that the appropriate $R$ to consider should be associated with the simulation's cell, $R=\left(3/4\pi\right)^{1/3}L_\mathrm{cell}$, where $L_\mathrm{cell}$ is the cell size of the low-resolution grid (for better precision, the realization of the perturbed box should be generated on a high-resolution grid). Note that the cell size cannot be arbitrarily small as we ought to have $M_\mathrm{min}<M_R$ for a non-zero local collapsed fraction (see square root in Eq.~\ref{eq: 23}). However, this condition is satisfied for a typical simulation cell of $R\sim L_\mathrm{cell}\sim1\,\mathrm{Mpc}$; according to Eq.~\eqref{eq: 24} such a smoothing radius corresponds to $M_R\sim10^{11}\,M_\odot$ and even in ACGs (whose associated $M_\mathrm{min}$ is much larger than in MCGs)~\cite{Barkana:2000fd}
\begin{flalign}\label{eq: 30}
\nonumber&M_{\mathrm{min}}^\mathrm{ACG}\left(z\right)\approx&
\\&\hspace{5mm}2.6\times10^{8}\,M_{\odot}\left(\frac{T_{\mathrm{vir}}}{1.98\times10^{4}\,\mathrm{K}}\frac{10}{1+z}\right)^{3/2}\left(\frac{h^{2}\Omega_{m}}{0.143}\right)^{-1/2},&
\end{flalign}
where $T_\mathrm{vir}\sim10^4\,\mathrm{K}$ is the typical temperature of a virialized halo of mass $M_\mathrm{min}$. Therefore, a simulation's cell of size $\sim1\,\mathrm{Mpc}$ contains enough mass to form stars in ACGs and MCGs. 

In contrast to the above, {\tt 21cmFAST} makes the following approximation. The radius $R$ in Eq.~\eqref{eq: 23} is associated with the comoving distance between $z$ and $z'$ ($z'>z$ is a dummy integration variable, denoting the redshift of the contributing sources whose radiation arrives to point $\mathbf x$ at redshift $z$), and no filtering process is done on the emissivity field\footnote{Furthermore, the inhomogeneous $M_\mathrm{min}^\mathrm{MCG}$ is also filtered and it is evaluated at $z$ rather at $z'$. This may cause a $\mathcal O\left(20\%\right)$  error in the power spectrum and was fixed in {\tt Zeus21}~\cite{Cruz:2024fsv}.}~\cite{Mesinger:2010ne} (see however Ref.~\cite{Reis:2021nqf} that correctly filters the emissivity field to account for the appropriate contributing sources).\footnote{Another approximation in {\tt 21cmFAST} and {\tt Zeus21} is the assumption of deterministic astrophysical relations to determine the emissivity field, but we note that Ref.~\cite{Nikolic:2024xxo} found that stochasticity is important in determining reionization history and the high redshift UV luminosity function.} Note that although {\tt Zeus21} does not work with a grid for speeding up the calculations, it was calibrated based on the output of {\tt 21cmFAST} and thus its output is indirectly affected by this approximation. Since the validity of this approximation was never tested, it is crucial to examine how the results of the simulation depend on it, a task which we leave for future work. In addition, we note that relaxing this approximation in {\tt 21cmFirstCLASS} would allow us to consider other window functions to account e.g.\ for multiple Ly$\alpha$ scattering, which has been argued to have a non-negligible impact on the 21-cm power spectrum~\cite{Reis:2021nqf, Semelin:2023lhz, Mittal:2023xih}.

\section{Conclusions}\label{sec: Conclusions}

Building a robust theoretical framework for the 21-cm signal is challenging, especially during the cosmic dawn and reionization epochs. Although valuable tools are available in the literature, all of them involve approximations or potential misinterpretations, often without thoroughly exploring their implications or consequences. For example, most of the codes model the 21-cm signal tracing the non-linear CDM density field, modeling its evolution using the SIGF. These choices glosses over a trivial but crucial aspect: since the 21-cm signal is sourced from neutral hydrogen atoms in the IGM, its modeling should follow the baryon density field. Even though baryons are coupled gravitationally to CDM at low redshifts, at higher redshifts this fact breaks down and the evolution of the two fields has to be carried on independently, also accounting for scale-dependent effects. Neglecting all of this potentially leads to misestimation of the 21-cm signal.

In this work, we compared two widely used public 21-cm codes, {\tt 21cmFAST}~\cite{Mesinger:2010ne, Munoz:2021psm} and {\tt Zeus21}~\cite{Munoz:2023kkg, Cruz:2024fsv}, with our code, {\tt 21cmFirstCLASS}~\cite{Flitter:2023mjj, Flitter:2023rzv}, an extension to {\tt 21cmFAST} that enables to begin the simulation from recombination with any desired cosmology. We outlined the main approximations introduced by the former and relaxed them in our implementation, demonstrating their impact on the 21-cm global signal and power spectrum.

We distinguished between baryons and CDM to carefully study the cosmological impact of the baryons on the 21-cm signal, similarly as in Ref.~\cite{Naoz:2005pd}, but we also extended our analysis to lower redshifts, after galaxies were formed. Unlike most studies which have adopted the SIGF approximation, whereby $\delta_b\propto D\left(z\right)$ and evolves in a scale-independent manner, we computed  the baryon density field more precisely in {\tt 21cmFirstCLASS} using the SDGF $\mathcal D_b\left(k,z\right)$. We have shown in previous work that this  method agrees very well with Boltzmann-Einstein solvers at the dark ages when all perturbations can be evolved linearly. Here, we extended this method to the regime when the Universe entered its non-linear phase, including cosmic dawn when galaxies and stars begun to form, as well as reionization.

We started our analysis from the dark ages, and demonstrated that the SIGF approximation fails poorly in that regime. We then moved to the non-linear epochs of cosmic dawn and reionization. As expected, even though some models of the SFRD consider the ``linear" $\delta_m^\mathrm{lin}$ as an input to the conditional HMF, it is inconsistent to consider the ``linear" $\delta_b^\mathrm{lin}$ once the dark ages are over, as this yields errors that are beyond the noise-level of 21-cm interferometers such as HERA. This is particularly true at low redshifts ($z\lesssim8$) and large scales ($k\lesssim0.3\,\mathrm{Mpc}^{-1}$), where the sensitivity of the interferometers is at maximum. Furthermore, we find that the SIGF approximation works very well for $z\lesssim12$, as this is the regime where the baryons and CDM can indeed be characterized as a single fluid. In contrast, at higher redshifts, the SIGF approximation may lead to $\mathcal O\left(10\%\right)$ error, implying that advanced detectors like SKA will be sensitive enough such that the SIGF approximation is inadequate for simulating the signal in the required precision, even during cosmic dawn.

In addition to understanding  how the 21-cm signal depends on the baryon over-density $\delta_b$, we also discussed the role of the total matter over-density $\delta_m$ in the calculation. This enters through the collapsed fraction, Eq.~\eqref{eq: 23}, as part of the EPS formalism. While it is clear that the 21-cm signal at cosmic dawn and reionization  should depend on the ``non-linear" $\delta_b$ (i.e., $\delta_b$ as evaluated by non-linear theory), two common approaches are considered in the literature for the total matter density field $\delta_m$. One option is to use the ``linear” $\delta_m^\mathrm{lin}$ since the EPS formalism is based on linear perturbation theory. Alternatively, one can choose to work with the non-linear $\delta_m^\mathrm{NL}$, as it was argued that this choice yields a better agreement with radiative transfer simulations. While {\tt Zeus21} ({\tt 21cmFAST}) adopts the former (latter) approach, {\tt 21cmFirtCLASS} enables the user to decide which $\delta_m$ should be used in Eq.~\eqref{eq: 23}. Regardless, as was discussed in the previous section, all codes still make in their calculations some assumptions whose validity should be further examined in the future.

In conclusion, we can summarize our detailed comparison between the three codes as follows (see also Fig.~\ref{fig: 7} and Table~\ref{tab: 1}). {\tt Zeus21} is a speedy code that attempts to reproduce the same 21-cm power spectrum as in {\tt 21cmFAST} (albeit on much shorter timescales), but it only works with linear theory for the density field. While it is debatable whether or not it is more appropriate to consider the linear matter density field $\delta_m^\mathrm{lin}$ in the EPS and SFRD calculations, it is clear that it is inconsistent to consider the linear baryon density field $\delta_b^\mathrm{lin}$ at low redshifts, where HERA and other 21-cm interferometers are most sensitive. Even though the exact separation between baryons and CDM is not critical for analyzing the data from current interferometers, it would become increasingly more important in the future when data from next generation interferometers like SKA becomes available. At the moment, {\tt 21cmFirstCLASS} is the only semi-numerical code that treats separately and consistently the different evolution in the baryon and CDM density fields during cosmic dawn and reionization epochs. Since {\tt 21cmFirstCLASS} is also capable of simulating the dark ages, where this effect is most pronounced, it is thus poised to serve as an important tool in analyzing the 21-cm signal as measured by future interferometers, whether they are ground-based or lunar-based.

\begin{acknowledgments}
It is our pleasure to thank Julian B. Mu\~noz for reviewing the manuscript and making comments that improved the quality of this paper. We wish to thank the anonymous referee for providing a thorough and insightful report. We would also like to thank Andrei Mesinger and Tal Adi for useful discussions. We also acknowledge the efforts of the  {\tt 21cmFAST} and {\tt CLASS} authors to produce state-of-the-art public 21-cm and CMB codes. JF is supported by the Zin fellowship awarded by the BGU Kreitmann School. SL is supported by an Azrieli International Postdoctoral Fellowship. EDK acknowledges  joint support from the U.S.-Israel Bi-national Science Foundation (BSF, grant No. 2022743) and  the U.S. National Science Foundation (NSF, grant No. 2307354), as well as support from the ISF-NSFC joint research program (grant No. 3156/23).
\end{acknowledgments}

\onecolumngrid

\appendix

\section{Boltzmann equation for the spin temperature}\label{sec: Boltzmann equation for the spin temperature}

Eq.~\eqref{eq: 3} is often used in the literature with the justification that the hyperfine transition rate is greater than the expansion rate of the Universe. This is indeed true during cosmic dawn and reionization epochs since 
\begin{equation}\label{eq: A1}
\frac{H\left(z\right)}{A_{10}}=4.3\times10^{-4}\left(\frac{h^2\Omega_m}{0.143}\right)^{1/2}\left(1+z\right)^{3/2}.
\end{equation}
However, above $z\gtrsim175$ this fraction becomes larger than unity, so it is unclear whether the equilibrium expression of Eq.~\eqref{eq: 3} is still valid at the dark ages. Indeed, Ref.~\cite{Lewis:2007kz} solved the fully relativistic Boltzmann Equation for the background spin temperature, while taking optical depth corrections to first order. In this subsection we quantify the validity of Eq.~\eqref{eq: 3}, while taking all orders in $\tau_{21}$.

We begin with the definition of the spin temperature from the fraction of hydrogen atoms that occupy the hyperfine triplet state, $n_1/n_0\equiv\left(g_1/g_0\right)\exp\left(-T_*/T_s\right)\approx3-3T_*/T_s$, where $g_1=3$ ($g_0=1$) is the total number of different triplet (singlet) states, and we assume $T_s\gg T_*$, a very good approximation when Eq.~\eqref{eq: 3} is considered (verified below).

The homogeneous Boltzmann equation for $n_0$ (i.e. with no collisional terms) can be determined from the number conservation of hydrogen nuclei $a^3n_\mathrm{H}=a^3n_\mathrm{HI}/\left(1-x_e\right)$, where $x_e\equiv n_e/n_\mathrm{H}=1-n_\mathrm{HI}/n_\mathrm{H}$ is the free electron fraction.$^*$\customfootnotetext{*}{For better comparison with Ref.~\cite{Lewis:2007kz}, in this appendix we use a slightly different definition for $x_e$ than the main text.}
\begin{equation}\label{eq: A2}
\frac{d}{dt}\left(a^3n_\mathrm{H}\right)=\frac{4a^3}{1-x_e}\left(\frac{dn_0}{dt}+3Hn_0+\frac{n_0}{1-x_e}\frac{dx_e}{dt}\right)=0,
\end{equation}
where we approximated $n_\mathrm{HI}\approx4n_0$. Without collisional terms, the right hand side 
is zero, while once we add them,
\begin{equation}\label{eq: A3}
\frac{dn_0}{dt}+3Hn_0+\frac{n_0}{1-x_e}\frac{dx_e}{dt}=-n_0\left(C_{01}+P_{01}+B_{01}I_\mathrm{CMB}\zeta_\mathrm{CMB}\right)+n_1\left(C_{10}+P_{10}+A_{10}+B_{10}I_\mathrm{CMB}\zeta_\mathrm{CMB}\right),
\end{equation}
where $B$'s are the Einstein coefficients for stimulated hyperfine transitions by CMB photons, and $C$'s ($P$'s) are the hyperfine excitation rate coefficients by particles collision (Ly$\alpha$ pumping). '01' labels correspond to hyperfine excitations, while '10' labels correspond to de-excitations to the ground state. $I_\mathrm{CMB}=2k_BT_\gamma/\lambda_{21}^2$ is the CMB black body radiation at the Rayleigh-Jeans limit near the 21-cm line, while $\zeta_\mathrm{CMB}$ accounts for its finite width.
This quantity can be related to the phase-space density in the vicinity of the 21-cm line, which we take to be~\cite{Venumadhav:2018uwn},
\begin{equation}\label{eq: A4}
f_{21}\left(\nu\right)=T_*^{-1}\left[T_s+\left(T_\gamma-T_s\right)\mathrm{e}^{-\tau_{21}\chi_{21}\left(\nu\right)}\right],
\end{equation}
where $\chi_{21}\left(\nu\right)$ is the cumulative function of the 21-cm line profile ($\chi_{21}$ goes from zero on the blue side to unity on the red side). Then, we find that
\begin{equation}\label{eq: A5}
\zeta_\mathrm{CMB}=\frac{T_*}{T_\gamma}\int_0^1d\chi_{21}\left(\nu\right)f_{21}\left(\nu\right)=1+\left(1-\frac{T_s}{T_\gamma}\right)\left(x_\mathrm{CMB}-1\right),
\end{equation}
where~\cite{Venumadhav:2018uwn}
\begin{equation}\label{eq: A6}
x_\mathrm{CMB}\equiv\frac{1-\mathrm{e}^{-\tau_{21}}}{\tau_{21}}.
\end{equation}
In most places in the literature, e.g.,~Ref.~\cite{Loeb:2003ya}, the zeroth order of the optical depth is considered with $x_\mathrm{CMB}=1$, and $\zeta_\mathrm{CMB}=1$. In Ref.~\cite{Lewis:2007kz} the first order is considered with $x_\mathrm{CMB}=1-\tau_{21}/2$ and $\zeta_\mathrm{CMB}=1-\left(1-T_s/T_\gamma\right)\tau_{21}/2$. Here, instead, we take all orders of the optical depth with $x_\mathrm{CMB}$ given by Eq.~\eqref{eq: A6}.

In order to proceed with the derivation, we shall assume detailed balance, implying that all sub-process are in equilibrium with their corresponding reverse sub-process. This implies the following relations,
\begin{eqnarray}
B_{10}=\frac{\lambda_{21}^3}{2\hbar c}A_{10},&\quad B_{01}&=\frac{g_1}{g_0}B_{10}=3B_{10},\quad\label{eq: A7} 
\\C_{01}=\frac{g_1}{g_0}C_{10}\mathrm{e}^{-T_*/T_k}\approx 3\left(1-\frac{T_*}{T_k}\right)C_{10},
&\quad&P_{01}=\frac{g_1}{g_0}P_{10}\mathrm{e}^{-T_*/T_k}\approx 3\left(1-\frac{T_*}{T_\alpha}\right)P_{10}.\label{eq: A8}
\end{eqnarray}
Inserting those relations in Eq.~\eqref{eq: A3} then yields
\begin{eqnarray}\label{eq: A9}
\nonumber\frac{1}{n_0}\frac{dn_0}{dt}+3H+\frac{1}{1-x_e}\frac{dx_e}{dt}=\left(\frac{n_1}{n_0}-3+3\frac{T_*}{T_k}\right)C_{10}+\left(\frac{n_1}{n_0}-3+3\frac{T_*}{T_\alpha}\right)P_{10}
\\+\left[\frac{n_1}{n_0}\left(1+\frac{T_\gamma}{T_*}\zeta_\mathrm{CMB}\right)-3\frac{T_\gamma}{T_*}\zeta_\mathrm{CMB}\right]A_{10}.
\end{eqnarray}

We would like to convert this equation to a differential equation for $T_s$. For this purpose, we compute the time derivative of $n_1/n_0$. Using again the fact that $d\left(a^3n_\mathrm{H}\right)/dt=0$, it is straightforward to show that
\begin{equation}\label{eq: A10}
\frac{d}{dt}\left(\frac{n_1}{n_0}\right)=-\left(\frac{1}{n_0}\frac{dn_0}{dt}+3H+\frac{1}{1-x_e}\frac{dx_e}{dt}\right)\left(1+\frac{n_1}{n_0}\right).
\end{equation}
On the other hand, since $n_1/n_0\equiv3\exp\left(-T_*/T_s\right)\approx3-3T_*/T_s$, then
\begin{equation}\label{eq: A11}
\frac{d}{dt}\left(\frac{n_1}{n_0}\right)=-3T_*\frac{dT_s^{-1}}{dt}.
\end{equation}
Combining Eqs.~\eqref{eq: A9}-\eqref{eq: A11} thus yields
\begin{equation}\label{eq: A12}
\frac{dT_s^{-1}}{dt}=4\left(T_k^{-1}-T_s^{-1}\right)C_{10}+4\left(T_\alpha^{-1}-T_s^{-1}\right)P_{10}+\frac{4}{T_*}\left(1-\frac{T_\gamma}{T_s}\zeta_\mathrm{CMB}\right)A_{10}.
\end{equation}
This is the Boltzmann equation for the (inverse) spin temperature. This equation is almost identical to Eq.~(28) in Ref.~\cite{Lewis:2007kz} with their approximation of $\zeta_\mathrm{CMB}\approx1-\left(1-T_s/T_\gamma\right)\tau_{21}/2$, but they have an extra $dx_e/dt$ term on the LHS which was canceled in our calculations.

\begin{SCfigure}
\includegraphics[width=0.5\columnwidth]{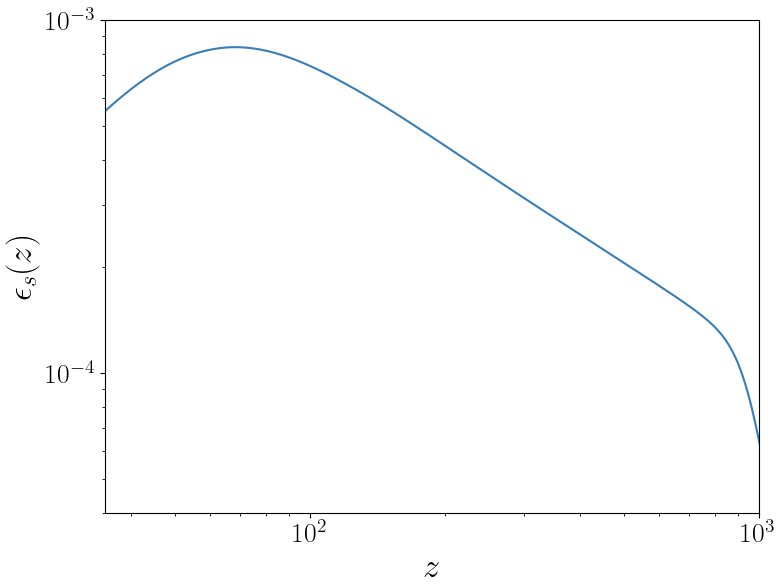}
\caption[0mm,5mm]{The inverse dimensionless rate $\epsilon_s\left(z\right)$ in the Boltzmann equation for the spin temperature, as given by Eq.~\eqref{eq: A14}. We assume that $\tilde x_\alpha=0$ during the dark ages. During cosmic dawn and reionization epochs, $\tilde x_\alpha\gg1$, thereby lowering the value of this quantity even further. Because $\epsilon_s\ll1$, we conclude that the equilibrium expression for $T_s$, Eq.~\eqref{eq: 13}, is valid also during the dark ages with excellent precision.}
\label{fig: 8}
\end{SCfigure}

To bring this equation to a more familiar form, we make use of $1-\left(T_\gamma/T_s\right)\zeta_\mathrm{CMB}=\left(1-T_\gamma/T_s\right)x_\mathrm{CMB}$ and we plug the dimensionless coupling coefficients, $P_{10}\equiv\left(T_\gamma A_{10}/T_*\right)\tilde x_\alpha$ and $C_{10}\equiv\left(T_\gamma A_{10}/T_*\right)x_\mathrm{coll}$. Finally, we also convert the time derivative on the LHS with redshift derivative via $dz/dt=-\left(1+z\right)H\left(z\right)$. We then arrive at
\begin{equation}\label{eq: A13}
\frac{dT_s^{-1}}{dz}=\frac{4T_\gamma A_{10}\left(x_\mathrm{CMB}+x_\mathrm{coll}+\tilde x_\alpha\right)}{T_*\left(1+z\right)H\left(z\right)}\left[T_s^{-1}-\frac{x_\mathrm{CMB}T_\gamma^{-1}+x_\mathrm{coll}T_k^{-1}+\tilde x_\alpha T_\alpha^{-1}}{x_\mathrm{CMB}+x_\mathrm{coll}+\tilde x_\alpha}\right].
\end{equation}
From here, we see that if
\begin{equation}\label{eq: A14}
\epsilon_s\left(z\right)\equiv\left[\frac{4T_\gamma A_{10}\left(x_\mathrm{CMB}+x_\mathrm{coll}+\tilde x_\alpha\right)}{T_*\left(1+z\right)H\left(z\right)}\right]^{-1}
\end{equation}
is much smaller than unity, then the following solution is an \emph{attractor},
\begin{equation}\label{eq: A15}
T_{s,\mathrm{eq}}^{-1}\equiv\frac{x_\mathrm{CMB}T_\gamma^{-1}+x_\mathrm{coll}T_k^{-1}+x_\alpha T_\alpha^{-1}}{x_\mathrm{CMB}+x_\mathrm{coll}+\tilde x_\alpha}.
\end{equation}
Then, under this condition ($\epsilon_s\ll1$) we can expand the solution of Eq.~\eqref{eq: A13} to first order in $\epsilon_s$, $T_s^{-1}=T_{s,\mathrm{eq}}^{-1}+\mathcal O\left(\epsilon_s\right)$. We plot $\epsilon_s\left(z\right)$ during the dark ages in Fig.~\ref{fig: 8}. We find that this quantity reaches maximum at $\epsilon_{s,\max}\left(z=69\right)\approx 10^{-3}$. Thus we estimate
\begin{equation}\label{eq: A16}
\frac{\left|T_s^{-1}-T_{s,\mathrm{eq}}^{-1}\right|}{T_s^{-1}}\lesssim \epsilon_{s,\max}=0.1\%.
\end{equation}
Hence, solving the Boltzmann equation as in Ref.~\cite{Lewis:2007kz} is unnecessary, and the equilibrium expression of Eq.~\eqref{eq: 13} holds also at the dark ages with excellent precision.

\section{Scale-dependent matter evolution}\label{sec: Scale-dependent matter evolution}

\begin{figure*}
\includegraphics[width=\textwidth]{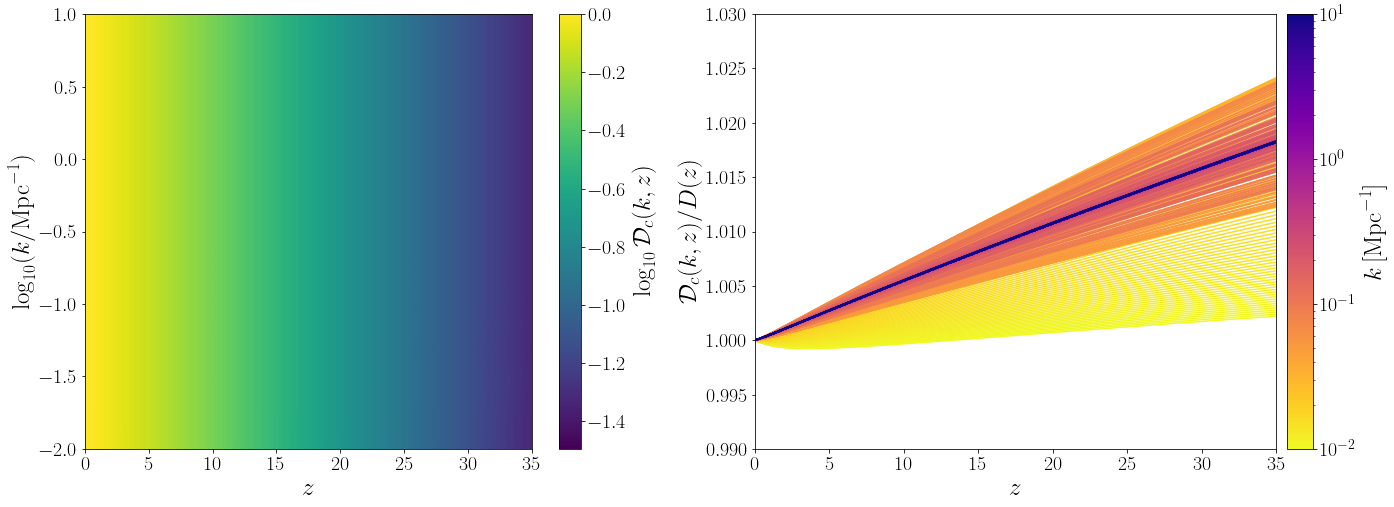}
\caption{Left panel: the scale-dependent growth factor for CDM. Right panel: comparison with the scale-independent growth factor. The oscillations with scale on the right panel indicate fake BAOs feature when the SIGF approximation is used.}
\label{fig: 9}
\end{figure*}

Even though CDM is considered as a collisionless fluid in $\Lambda$CDM, the evolution of its density fluctuations is not perfectly scale invariant, due to the gravitational coupling with baryons. Indeed, at low redshifts ($z\lesssim 1$) $\delta_c$ and $\delta_b$ approach each other, and BAO features from $\delta_b$ can be seen in the CDM transfer function $\mathcal T_c\left(k,z\right)$. This is no longer true at the high redshifts relevant for the 21-cm signal. Thus, the commonly used linear approximation of $\delta_c\left(\mathbf k,z\right)\approx D\left(z\right)\delta_0\left(\mathbf k\right)$ results in fake BAO features in $\delta_c\left(\mathbf k,z\right)$ at high redshifts ($z\gtrsim 1$). This can be remedied by using the appropriate SDGF for CDM $\delta_c\left(\mathbf k,z\right)\approx \mathcal D_c\left(k,z\right)\delta_0\left(\mathbf k\right)$, where $\mathcal D_c\left(k,z\right)$ is defined in Eq.~\eqref{eq: 14}. We compare between $\mathcal D_c\left(k,z\right)$ and $D\left(z\right)$ in Fig.~\ref{fig: 9}. From the left panel, it indeed appears as if $\mathcal D_c\left(k,z\right)$ is scale-invariant, unlike $\mathcal D_b\left(k,z\right)$ (c.f.~Fig.~\ref{fig: 1}). However, a closer comparison between $\mathcal D_c\left(k,z\right)$ and $D\left(z\right)$ on the right panel shows that the two can differ by more than 2\% at high redshifts ($z\gtrsim25$), especially at small scales ($k\gtrsim1\,\mathrm{Mpc}^{-1}$). Moreover, the fake BAO that the SIGF approximation yields is clearly seen on the right panel, through the non-monotonous behavior of $\mathcal D_c\left(k,z\right)/D\left(z\right)$ when $k$ is increased.

As we discussed in Sec.~\ref{subsec: Star formation rate density}, the local SFRD can be computed with the EPS formalism, where the local linear matter over-density $\delta_m^\mathrm{lin}\left(\mathbf x,z\right)$ is considered. This quantity can be computed in Fourier space via $\delta^\mathrm{lin}_m\left(\mathbf k,z\right)\approx \mathcal D_m\left(k,z\right)\delta_0\left(\mathbf k\right)$, where
\begin{equation}\label{eq: B1}
\mathcal D_m\left(k,z\right)\equiv\frac{\mathcal T_m\left(k,z\right)}{\mathcal T_m\left(k,z=0\right)}\approx\frac{\Omega_b}{\Omega_m}\frac{\mathcal T_b\left(k,z\right)}{\mathcal T_b\left(k,z=0\right)}+\frac{\Omega_c}{\Omega_m}\frac{\mathcal T_c\left(k,z\right)}{\mathcal T_c\left(k,z=0\right)}=\frac{\Omega_b}{\Omega_m}\mathcal D_b\left(k,z\right)+\frac{\Omega_c}{\Omega_m}\mathcal D_c\left(k,z\right),
\end{equation}
where we used the excellent approximation that $\mathcal T_b$, $\mathcal T_c$ and $\mathcal T_m$ are all the same at $z=0$. This definition allows us to write $\sigma^2\left(M_h,z\right)$ from Eq.~\eqref{eq: 20} as
\begin{equation}\label{eq: B2}
\sigma^2\left(M_h,z\right)=\int_0^\infty\frac{dk}{k} A_s\left(\frac{k}{k_\star}\right)^{n_s-1}\mathcal T^2_m\left(k,z=0\right)\mathcal D_m^2\left(k,z\right)W^2\left(kR_{M_h}\right).
\end{equation}

\begin{figure*}
\includegraphics[width=0.86\textwidth]{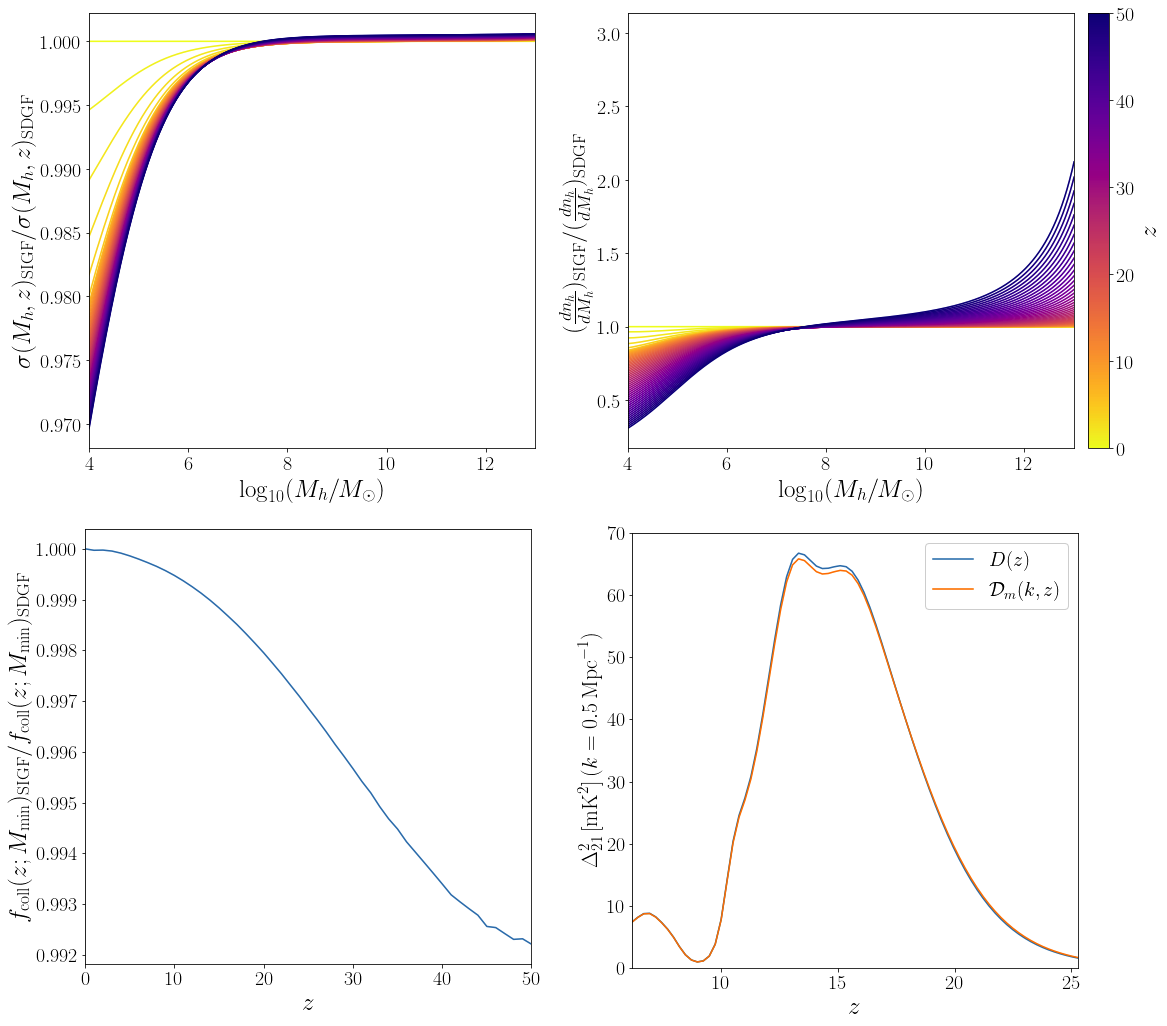}
\caption{Top panels: The ratio of $\sigma\left(M_h,z\right)$ and the Sheth-Tormen HMF, when they are calculated with the SIGF and the SDGF. Bottom left panel: The ratio of the collapsed fraction of halos that host ACGs (with $M_\mathrm{min}$ given by Eq.~\ref{eq: 30}), when calculated with the SIGF and the SDGF. Bottom right panel: The 21-cm power spectrum as a function of redshift at $k=0.5\,\mathrm{Mpc}^{-1}$, when the matter over-density field is evolved with the SIGF (blue curve) or the SDGF (orange curve).}
\label{fig: 10}
\end{figure*}

\newpage
In the SIGF approximation, it is assumed that $\mathcal D_m\left(k,z\right)\approx D\left(z\right)$ and hence $\sigma\left(M_h,z\right)\propto D\left(z\right)$. Therefore, the fake BAOs induced by the SIGF approximation at high redshifts cause an error of a few percent in $\sigma\left(M_h,z\right)$, mainly in low halo masses, and an order unity error in the derived HMF, as we show in the top panels of Fig.~\ref{fig: 10}. In making the comparison for the HMF though, it is important to bare in mind that the ST parameters were fitted to the HMF that is computed with the SIGF approximation. It is thus intriguing to wonder if the output from N-body simulations can be better fitted by relaxing the SIGF approximation.

When the collapsed fraction in ACGs is considered, the lower limit of the integral in Eq.~\eqref{eq: 21} is at $M_h\sim10^8\,M_\odot$, which is coincidently the regime where the error in the HMF is minimum, while higher halo masses less contribute to the integral. Thus, the error in the collapsed fraction of halos that host ACGs is sub-percent, as shown in the bottom left panel in Fig.~\ref{fig: 10}. When MCGs are considered too, their $M_\mathrm{min}$ is smaller than the equivalent quantity in ACGs by an order of magnitude, and it is subject to inhomogeneous baryonic feedback such as LW radiation and relative velocity with respect to CDM. We compare in the last panel of Fig.~\ref{fig: 10} how the SIGF approximation, applied both in ACGs and MCGs, affects the 21-cm power spectrum. As it appears, the error introduced by the SIGF approximation in the HMF at high redshifts has almost no effect on the 21-cm power spectrum. This conclusion however relies on the assumed astrophysical model and the fiducial astrophysical parameters we have chosen. In other models, where the star production rate is predicted to be more efficient at high redshifts, the SIGF approximation may yield larger errors in the 21-cm signal. We defer this investigation for future work.

\section{Scale-dependent Lagrangian perturbation theory}\label{sec: Scale-dependent Lagrangian perturbation theory}

Normally, in Lagrangian perturbation theory~\cite{1978ApJ...221..937F, Buchert:1989xx, Moutarde:1991evx, 1992MNRAS.254..729B, 1993A&A...267L..51B, Bouchet:1994xp, Hivon:1994qb, Buchert:1995bq, Tatekawa:2004mq, Matsubara:2008wx, Carlson:2009it, Zheligovsky:2013eca}, the SIGF approximation is taken, namely treating the baryon and CDM density fields the same and solving for the total matter density $\delta_m$. In what follows, we relax this approximation and treat the baryons and CDM as two distinguished fluids. In the Lagrangian frame (where we go with the flow, i.e.~we follow parcels along their trajectory), the equations of motion are
\begin{eqnarray}
\dot\delta_i+a^{-1}\boldsymbol\nabla_{\mathbf x}\cdot\mathbf v_i&=&0\label{eq: C1}
\\\dot{\mathbf v}_i+H\mathbf v_i+a^{-1}\boldsymbol\nabla_{\mathbf x}\Phi&=&0\label{eq: C2}
\end{eqnarray}
where $i=b,c$ for baryons and CDM (to reduce clutter, we omit the $i$ index when we write derivatives with respect to space). Note that we have ignored the sound speed term in the Euler equation as we focus on super Jeans scales. In addition, we have the Poisson equation
\begin{equation}
\nabla^2\Phi_{\mathbf x}=4\pi G\bar\rho_m a^2\delta_m=4\pi G\bar\rho_m a^2\sum_j\frac{\Omega_j}{\Omega_m}\delta_j.\label{eq: C3}
\end{equation}

In LPT, for each species $i$, we relate the initial Lagrangian coordinates $\mathbf q^{(i)}$ to the Eulerian coordinates $\mathbf x^{(i)}$ via
\begin{equation}\label{eq: C4}
\mathbf x^{(i)}\left(\mathbf q^{(i)},z\right)=\mathbf q^{(i)}\left(z_0\right)+\mathbf s^{(i)}\left(\mathbf q^{(i)},z\right),
\end{equation}
where $\mathbf s^{(i)}$ is the displacement field vector for species $i$ and $z_0\gg z$ is a high redshift in which the initial conditions for the calculation are given. From here, we see that $\mathbf v_i\equiv a\dot{\mathbf x}^{(i)}=a\dot{\mathbf s}^{(i)}$. We now take the divergence of the Euler equation (Eq.~\eqref{eq: C2}), and substitute the Poisson equation (Eq.~\eqref{eq: C3}) to write
\begin{equation}\label{eq: C5}
\boldsymbol\nabla_{\mathbf x}\cdot\left(\ddot{\mathbf s}^{(i)}+H\dot{\mathbf s}^{(i)}\right)=-4\pi G\bar\rho_m\sum_j\frac{\Omega_j}{\Omega_m}\delta_j.
\end{equation}
We shall focus on irrotational trajectories that obey $\boldsymbol\nabla_{\mathbf q}\times\mathbf s^{(i)}=0$, implying that $\mathbf s^{(i)}$ can be expressed as the gradient of a scalar field $S_i$,
\begin{equation}\label{eq: C6}
\mathbf s^{(i)}\left(\mathbf q,z\right)\equiv\boldsymbol\nabla_{\mathbf q}S_i\left(\mathbf q,z\right)-\boldsymbol\nabla_{\mathbf q}S_i\left(\mathbf q,z_0\right).
\end{equation}
Note that by construction, $\mathbf{s}^{(i)}\left(\mathbf{q},z_{0}\right)\equiv0$, i.e we impose the initial conditions $\mathbf{x}^{(i)}\left(z_{0}\right)\equiv\mathbf{q}^{(i)}\left(z_{0}\right)$.

The continuity equation implies mass conservation for each species $j$, $\rho_j\left(\mathbf x\right)d^3x=\rho_j\left(\mathbf q\right)d^3q$ (here $\rho$ denotes the comoving energy density, without the $\left(1+z\right)^3$ scaling). We can use the Jacobian $J_j\equiv\det\left(\partial x_n^{(j)}/\partial q_m\right)$ to write
\begin{equation}\label{eq: C7}
\rho_j\left(\mathbf q,z\right)J_jd^3q=\rho_j\left(\mathbf q,z_0\right)d^3q\qquad\implies\qquad J_j=\frac{\rho_j\left(\mathbf q,z_0\right)}{\rho_j\left(\mathbf q,z\right)}\approx\left(1+\delta_j\right)^{-1}
\end{equation}
where the last equality assumes that $\rho_j\left(\mathbf q, z\right)=\bar\rho_j$, i.e. that the initial density field is nearly unperturbed at $z=z_0$. This assumption can be relaxed, although the approximation $J_j\approx\left(1+\delta_j\right)^{-1}$ will allow us to write Eq.~\eqref{eq: C11} without $S_i\left(\mathbf q,z_0\right)$ correction terms. At the end, keeping $S_i\left(\mathbf q,z_0\right)$ and $\delta_i\left(\mathbf q,z_0\right)$ yields the same solutions we derive, which is why we omit them for brevity. Combining Eqs.~\eqref{eq: C5}, \eqref{eq: C6} and \eqref{eq: C7} then gives
\begin{equation}\label{eq: C8}
\boldsymbol\nabla_{\mathbf{x}}\cdot\boldsymbol\nabla_{\mathbf{q}}\left(\ddot{S}_{i}+2H\dot{S}_{i}\right)=-4\pi G\bar{\rho}_m\sum_{j}\frac{\Omega_{j}}{\Omega_{m}}\left(J_{j}^{-1}-1\right).
\end{equation}
This is one equation we have. We also have an equation that relates the gradients (Einstein summation convention is assumed for the dummy index $m$),
\begin{equation}\label{eq: C9}
\frac{\partial}{\partial q_n}=\frac{\partial}{\partial x_n}+\frac{\partial s_{m}^{(i)}}{\partial q_{n}}\frac{\partial}{\partial x_{m}}=\frac{\partial}{\partial x_n}+\frac{\partial^{2}S_{i}}{\partial q_{n}\partial q_{m}}\frac{\partial}{\partial x_{m}}.
\end{equation}
Finally, we have an equation that relates $J_{j}$ to $S_{j}$. Since for any $3\times3$ matrix $\mathbf M$, we have $\det \mathbf M=\epsilon_{ijk}\epsilon_{abc}M_{ia}M_{jb}M_{kc}$, it is straightforward to show that
\begin{equation}\label{eq: C10}
J_{j}\equiv\det\left(\frac{\partial x_{n}^{(j)}}{\partial q_{m}}\right)=\det\left(\delta_{nm}+\frac{\partial s_{n}^{(j)}}{\partial q_{m}}\right)=1+\boldsymbol\nabla_{\mathbf{q}}\cdot\mathbf{s}^{(j)}+\frac{1}{2}\left[\left(\boldsymbol\nabla_{\mathbf{q}}\cdot\mathbf{s}^{(j)}\right)^{2}-\frac{\partial s_{n}^{(j)}}{\partial q_{m}}\frac{\partial s_{m}^{(j)}}{\partial q_{n}}\right]+\det\left(\frac{\partial s_{n}^{(j)}}{\partial q_{m}}\right),
\end{equation}
or,
\begin{equation}\label{eq: C11}
J_{j}=1+\nabla_{\mathbf{q}}^{2}S_{j}+\frac{1}{2}\left[\left(\nabla_{\mathbf{q}}^{2}S_{j}\right)^{2}-\left(\frac{\partial^{2}S_{j}}{\partial q_{n}\partial q_{m}}\right)^{2}\right]+\det\left(\frac{\partial^{2}S_{j}}{\partial q_{n}\partial q_{m}}\right).
\end{equation}
Equations \eqref{eq: C8}, \eqref{eq: C9}, \eqref{eq: C11} form a set of non-linear equations for $S_{i}$. We use perturbative approach to solve them.

\subsection{First order - Zel'dovich approximation}\label{subsec: First order - Zel'dovich approximation}

Let us expand $S_{i}\left(\mathbf{q},z\right)$ to its first order, $S_{i}\left(\mathbf{q},z\right)\approx S_{i}^{(1)}\left(\mathbf{q},z\right)$. Then, to first order,
\begin{equation}\label{eq: C12}
J_{j}\approx1+\nabla_{\mathbf{q}}^{2}S_{j}^{\left(1\right)}\implies\delta_{j}=J_{j}^{-1}-1=\left(1+\nabla_{\mathbf{q}}^{2}S_{j}^{\left(1\right)}\right)^{-1}-1\approx-\nabla_{\mathbf{q}}^{2}S_{j}^{\left(1\right)}.
\end{equation}
From here, we see that in Fourier space,
\begin{equation}\label{eq: C13}
S_{j}^{\left(1\right)}\left(\mathbf{k},z\right)=\frac{\delta_{j}\left(\mathbf{k},z\right)}{k^{2}}\approx\mathcal{D}_{j}\left(k,z\right)\frac{\delta_{0}\left(\mathbf{k}\right)}{k^{2}},
\end{equation}
where we used the linear expression for $\delta_{i}$ (Eq.~\ref{eq: 13}) as we seek first order solution to $S_{i}$. To see that this solution is consistent, we expand the Euler-Poisson equation, Eq.~\eqref{eq: C8}, to first order,
\begin{equation}\label{eq: C14}
\nabla_{\mathbf{q}}^{2}\left(\ddot{S}_{i}^{\left(1\right)}\left(\mathbf{q},z\right)+2H\dot{S}_{i}^{\left(1\right)}\left(\mathbf{q},z\right)-4\pi G\bar{\rho}_m\sum_{j}\frac{\Omega_{j}}{\Omega_{m}}S_{j}^{\left(1\right)}\left(\mathbf{q},z\right)\right)=0.
\end{equation}
Then, by plugging $\delta_{i}\left(\mathbf{q},z\right)=-\nabla_{\mathbf{q}}^{2}S_{i}^{\left(1\right)}\left(\mathbf{q},z\right)$, we find
\begin{equation}\label{eq: C15}
\ddot{\delta}_{i}\left(\mathbf{q},z\right)+2H\dot{\delta}_{i}\left(\mathbf{q},z\right)-4\pi G\bar{\rho}_m\sum_{j}\frac{\Omega_{j}}{\Omega_{m}}\delta_{j}\left(\mathbf{q},z\right)=0,
\end{equation}
which is indeed the equation of motion for $\delta_i$ in linear perturbation theory, above the Jeans scale (c.f.~Eq.~\ref{eq: 11}). Furthermore, the homogeneity assumption we made in Eq.~\eqref{eq: C7} implies that $\delta_i$ and $S_i^{(1)}$ have the same zero initial conditions at $z=z_0$. Thus, the solution to $\mathbf{s}_{i}^{\left(1\right)}$ in Fourier space reads
\begin{equation}\label{eq: C16}
\mathbf{s}_{i}^{\left(1\right)}\left(\mathbf{k},z\right)=\left[\mathcal{D}_{i}\left(k,z\right)-\mathcal{D}_{i}\left(k,z_{0}\right)\right]\frac{i\mathbf{k}}{k^{2}}\delta_{0}\left(\mathbf{k}\right).
\end{equation}
This equation is consistent with our ansatz in Ref.~\cite{Flitter:2023rzv} that was based on the Zel'dovich approximation~\cite{1970A&A.....5...84Z, White:2014gfa, Hidding:2013kka}. Note that given the SDGF, the solution for $\mathbf{s}_{i}^{\left(1\right)}\left(\mathbf{k},z\right)$ is independent of its history, e.g.~we can find $\mathbf{s}_{i}^{\left(1\right)}\left(\mathbf{k},z=20\right)$ before we solve for $\mathbf{s}_{i}^{\left(1\right)}\left(\mathbf{k},z=30\right)$. In addition, note that $\mathbf{s}_{i}^{\left(1\right)}$ cannot be separated to its spatial and temporal components, as this separation cannot be done for the SDGF $\mathcal D_i\left(k,z\right)$. However, this becomes possible when replacing the SDGF with the SIGF $\mathcal D_i\left(k,z\right)\to D\left(z\right)$,
\begin{equation}\label{eq: C17}
\mathbf{s}_{i}^{\left(1\right)}\left(\mathbf{k},z\right)\approx\left[D\left(z\right)-D\left(z_{0}\right)\right]\frac{i\mathbf{k}}{k^{2}}\delta_{0}\left(\mathbf{k}\right),
\end{equation}
as commonly done in the literature~\cite{1985ApJS...57..241E, Sirko:2005uz, Mesinger:2007pd}.

\subsection{Second order - 2LPT}\label{subsec: Second order - 2LPT}

Next, we expand $S_{i}$ to its second order, $S_{i}\left(\mathbf{q},z\right)\approx S_{i}^{(1)}\left(\mathbf{q},z\right)+S_{i}^{(2)}\left(\mathbf{q},z\right)$. The second order expansion of the Jacobian is
\begin{equation}\label{eq: C18}
J_j\approx 1+\nabla_{\mathbf{q}}^{2}\left(S_{j}^{\left(1\right)}+S_{j}^{\left(2\right)}\right)+\frac{1}{2}\left[\left(\nabla_{\mathbf{q}}^{2}S_{j}^{\left(1\right)}\right)^{2}-\left(\frac{\partial^{2}S_{j}^{\left(1\right)}}{\partial q_{n}\partial q_{m}}\right)^{2}\right].
\end{equation}
Therefore, the second order expansion of the RHS of Eq.~\eqref{eq: C8} is
\begin{equation}\label{eq: C19}
-4\pi G\bar{\rho}\sum_{j}\frac{\Omega_{j}}{\Omega_{m}}\left(J_{j}^{-1}-1\right)\approx 4\pi G\bar{\rho}_{m}\sum_{j}\frac{\Omega_{j}}{\Omega_{m}}\left\{ \nabla_{\mathbf{q}}^{2}S_{j}^{\left(1\right)}+\nabla_{\mathbf{q}}^{2}S_{j}^{\left(2\right)}-\frac{1}{2}\left[\left(\nabla_{\mathbf{q}}^{2}S_{j}^{\left(1\right)}\right)^{2}+\left(\frac{\partial^{2}S_{j}^{\left(1\right)}}{\partial q_{n}\partial q_{m}}\right)^{2}\right]\right\},
\end{equation}
while the second order expansion of the LHS is
\begin{equation}\label{eq: C20}
\nabla_{\mathbf{x}}\cdot\nabla_{\mathbf{q}}\left(\ddot{S}_{i}+2H\dot{S}_{i}\right)\approx \nabla_{\mathbf{q}}^{2}\left(\ddot{S}_{i}^{\left(1\right)}+\ddot{S}_{i}^{\left(2\right)}+2H\dot{S}_{i}^{\left(1\right)}+2H\dot{S}_{i}^{\left(2\right)}\right)-\frac{\partial^{2}S_{i}^{\left(1\right)}}{\partial q_{n}\partial q_{m}}\frac{\partial^2}{\partial q_{n}q_{m}}\left(\ddot{S}_{i}^{\left(1\right)}+2H\dot{S}_{i}^{\left(1\right)}\right).
\end{equation}
Combining Eqs.~\eqref{eq: C19}-\eqref{eq: C20}, together with the differential equation we obtained for $S^{(1)}$, Eq.~\eqref{eq: C14}, yields
\begin{flalign}\label{eq: C21}
&\nonumber\nabla_{\mathbf{q}}^{2}\left(\ddot{S}_{i}^{\left(2\right)}+2H\dot{S}_{i}^{\left(2\right)}-4\pi G\bar{\rho}_{m}\sum_{j}\frac{\Omega_{j}}{\Omega_{m}}S_{j}^{\left(2\right)}\right)=&
\\&\hspace{20mm}\frac{\partial^{2}S_{i}^{\left(1\right)}}{\partial q_{n}\partial q_{m}}\frac{\partial^2}{\partial q_{n}q_{m}}\left(\ddot{S}_{i}^{\left(1\right)}+2H\dot{S}_{i}^{\left(1\right)}\right)-2\pi G\bar{\rho}_{m}\sum_{j}\frac{\Omega_{j}}{\Omega_{m}}\left[\left(\nabla_{\mathbf{q}}^{2}S_{j}^{\left(1\right)}\right)^{2}+\left(\frac{\partial^{2}S_{j}^{\left(1\right)}}{\partial q_{n}\partial q_{m}}\right)^{2}\right].&
\end{flalign}
If variable separation was possible, then from Eq.~\eqref{eq: C21} we could have obtained two different equations, one for the spatial part of $S_i^{(2)}$, and one for its temporal part. However, as we established in Eq.~\eqref{eq: C13}, variable separation cannot be performed for $S_i^{(1)}$, and thus $S_i^{(2)}$ cannot be separated to its spatial and temporal components either. 

Note that given $S_i^{(1)}$, Eq.~\eqref{eq: C21} does not give a differential equation for $S_i^{(2)}$, but rather for the Laplacian of $S_i^{(2)}$. In order to obtain an integro-differential equation for $S_i^{(2)}$, we take the Fourier transform of Eq.~\eqref{eq: C21}. While applying Fourier transform on the LHS gives a multiplication by $-k^2$, taking the Fourier transform of the RHS, being proportional to the square of second derivatives of $S_i^{(1)}$, would result in a convolution. We use the fact that in Fourier space $\ddot{S}_{i}^{\left(1\right)}\left(\mathbf{k},z\right)+2H\dot{S}_{i}^{\left(1\right)}\left(\mathbf{k},z\right)=4\pi G\bar{\rho}_m\sum_{j}\frac{\Omega_{j}}{\Omega_{m}}S_{j}^{\left(1\right)}\left(\mathbf{k},z\right)$ and apply the inverse Fourier transform to arrive at
\begin{flalign}\label{eq: C22}
\nonumber&\ddot{S}_{i}^{\left(2\right)}\left(\mathbf{q},z\right)+2H\dot{S}_{i}^{\left(2\right)}\left(\mathbf{q},z\right)-4\pi G\bar{\rho}_{m}\sum_{j}\frac{\Omega_{j}}{\Omega_{m}}S_{j}^{\left(2\right)}\left(\mathbf{q},z\right)&
\\&=-4\pi G\bar{\rho}_{m}\sum_{j}\frac{\Omega_{j}}{\Omega_{m}}\int\frac{d^{3}k}{\left(2\pi\right)^{3}}\int\frac{d^{3}k'}{\left(2\pi\right)^{3}}\left[\left(\mathbf{k}\cdot\mathbf{k}'\right)^{2}S_{i}^{\left(1\right)}\left(\mathbf{k}',z\right)-\frac{1}{2}\left[k^{2}k'^{2}+\left(\mathbf{k}\cdot\mathbf{k}'\right)^{2}\right]S_{j}^{\left(1\right)}\left(\mathbf{k}',z\right)\right]\frac{S_{j}^{\left(1\right)}\left(\mathbf{k},z\right)}{\left|\mathbf{k}+\mathbf{k}'\right|^{2}}\mathrm{e}^{i\left(\mathbf{k}+\mathbf{k'}\right)\cdot\mathbf{q}}.&
\end{flalign}
This integro-differential equation actually contains two equations, one for the baryons ($i=b$) and one for CDM ($i=c$), and these two equations have to be simultaneously solved due to the gravitational coupling between the species, while imposing zero initial conditions for $S_i^{\left(2\right)}$ and its derivative at $z=z_0$.

\subsubsection{Scale independent growth}\label{subsubsec: Scale independent growth}

While in principle Eq.~\eqref{eq: C22} can be solved numerically, this is computationally expensive, as for each cell $\mathbf q$ in the grid, double integration (that involves all the cells in the box in Fourier space) is required, for both species. To ease the computational cost, we will make the approximation that is done vastly in the literature, of approximating $\mathcal D_i\left(k,z\right)\approx D\left(z\right)$, i.e.,~we replace the SDGF for each species with a universal SIGF. Then, from Eq.~\eqref{eq: C13}, we write
\begin{equation}\label{eq: C23}
S_{i}^{\left(1\right)}\left(\mathbf{q},z\right)\approx D\left(z\right)\psi^{\left(1\right)}\left(\mathbf{q}\right)\equiv S^{\left(1\right)}\left(\mathbf{q},z\right),
\end{equation}
where $\psi^{\left(1\right)}\left(\mathbf{q}\right)$ is the inverse Fourier transform of $\delta_0\left(\mathbf k\right)/k^2$ (or equivalently, in real space $\delta_0=-\nabla^2_{\mathbf q}\psi^{(1)}$). Because $\psi^{(1)}$ is derived from $\delta_0$, which is common for all species, under the SIGF approximation the solution to $S_{i}^{\left(1\right)}$ is a universal (i.e.~independent of the species $i$) solution $S^{\left(1\right)}$. 
Plugging this universal solution back to Eq.~\eqref{eq: C22} yields
\begin{flalign}\label{eq: C24}
\nonumber&\ddot{S}_{i}^{\left(2\right)}\left(\mathbf{q},z\right)+2H\dot{S}_{i}^{\left(2\right)}\left(\mathbf{q},z\right)-4\pi G\bar{\rho}_{m}\sum_{j}\frac{\Omega_{j}}{\Omega_{m}}S_{j}^{\left(2\right)}\left(\mathbf{q},z\right)&
\\&\hspace{20mm}=4\pi G\bar{\rho}_{m}D\left(z\right)\int\frac{d^{3}k}{\left(2\pi\right)^{3}}\int\frac{d^{3}k'}{\left(2\pi\right)^{3}}\left[\frac{1}{2}k^{2}k'^{2}-\frac{1}{2}\left(\mathbf{k}\cdot\mathbf{k}'\right)^{2}\right]\frac{\psi^{\left(1\right)}\left(\mathbf{k}\right)\psi^{\left(1\right)}\left(\mathbf{k}'\right)}{\left|\mathbf{k}+\mathbf{k}'\right|^{2}}\mathrm{e}^{i\left(\mathbf{k}+\mathbf{k'}\right)\cdot\mathbf{q}}.&
\end{flalign}
Because the RHS of Eq.~\eqref{eq: C24} is universal for all species, and since $S_i^{(2)}$ has zero initial conditions at $z=z_0$ for every species $i$, we deduce that the SIGF approximation implies that $S_i^{(2)}\left(\mathbf q,z\right)$ also has a universal solution $S^{(2)}\left(\mathbf q,z\right)$. This property can be further inferred for any higher order in perturbation theory. Hence, the SIGF approximation corresponds to solving the perturbed equations of motions under the approximation $\delta_b\approx\delta_c\approx\delta_m$, namely we directly solve for the total matter density field.

After substituting the universal solution $S_i^{(2)}\left(\mathbf q,z\right)=S^{(2)}\left(\mathbf q,z\right)$, Eq.~\eqref{eq: C24} can be written in the following form,
\begin{equation}\label{eq: C25}
\nabla^2_{\mathbf q}\left[\ddot{S}^{\left(2\right)}\left(\mathbf{q},z\right)+2H\dot{S}^{\left(2\right)}\left(\mathbf{q},z\right)-4\pi G\bar{\rho}_{m}S^{\left(2\right)}\left(\mathbf{q},z\right)\right]=-4\pi G\bar{\rho}_{m}D^2\left(z\right)\left[\frac{1}{2}\left(\nabla_{\mathbf{q}}^{2}\psi^{\left(1\right)}\right)^{2}-\frac{1}{2}\left(\frac{\partial^{2}\psi^{\left(1\right)}}{\partial q_{n}\partial q_{m}}\right)^{2}\right].
\end{equation}
Because of the variable separation on the RHS, we see that $S^{(2)}$ can be decomposed to its spatial and temporal parts
\begin{equation}\label{eq: C26}
S^{\left(2\right)}\left(\mathbf{q},z\right)=E\left(z\right)\psi^{\left(2\right)}\left(\mathbf{q}\right),
\end{equation}
where $E\left(z\right)$ and $\psi^{\left(2\right)}\left(\mathbf{q}\right)$ obey the following differential equations,
\begin{equation}\label{eq: C27}
\ddot{E}+2H\dot{E}-4\pi G\bar{\rho}_mE=-4\pi G\bar{\rho}_{m}D^{2}
\end{equation}
\begin{equation}\label{eq: C28}
\nabla_{\mathbf{q}}^{2}\psi^{\left(2\right)}=\frac{1}{2}\left[\left(\nabla_{\mathbf{q}}^{2}\psi^{\left(1\right)}\right)^{2}-\left(\frac{\partial^{2}\psi^{\left(1\right)}}{\partial q_{n}\partial q_{m}}\right)^{2}\right].
\end{equation}
As we show below, the solution to $E$ is $E\approx-{3}D^{2}/{7}$~\cite{Bouchet:1994xp, Scoccimarro:1997gr}, and thus the second order correction to the displacement vector is
\begin{equation}\label{eq: C29}
\mathbf{s}^{\left(2\right)}\left(\mathbf{q},z\right)=-\frac{3}{7}\left[D^{2}\left(z\right)-D^{2}\left(z_{0}\right)\right]\boldsymbol\nabla_{\mathbf{q}}\psi^{\left(2\right)}\left(\mathbf{q}\right).
\end{equation}
This closed solution for the displacement vector allows one to compute $\mathbf{s}^{\left(2\right)}$ for any $z$, independently from its history. 

\subsubsection{Solution to $E\left(z\right)$}\label{subsubsec: Solution to E}

Before solving for $E\left(z\right)$, we shall solve first the differential equation for the SIGF $D\left(z\right)$. This equation is obtained from Eq.~\eqref{eq: C15} when substituting $\delta_i\left(\mathbf q,z\right)=D\left(z\right)\delta_0\left(\mathbf q\right)$,
\begin{equation}\label{eq: C30}
\ddot D+2H\dot D-4\pi G\bar{\rho}_mD=0.
\end{equation}
During matter domination, we know that the scale factor evolves in time as $a\propto t^{2/3}$ and thus $H\equiv\dot a/a=\left(2/3\right)t^{-1/3}$. We also know, from the Friedmann equation, that
\begin{equation}\label{eq: C31}
4\pi G\bar{\rho}_{m}=\frac{3}{2}H^{2}\Omega_{m}\approx\frac{2}{3}t^{-2}.
\end{equation}
where we assumed for simplicity $\Omega_{m}=1$. Thus the equation of motion for the SIGF becomes
\begin{equation}\label{eq: C32}
\ddot{D}+\frac{4}{3}t^{-1}\dot{D}-\frac{2}{3}t^{-2}D=0.
\end{equation}
To solve this homogeneous differential equation, we guess $D\left(t\right)\propto t^{n}$ and plug our guess back in Eq.~\eqref{eq: C32}, 
\begin{equation}\label{eq: C33}
n\left(n-1\right)+\frac{4}{3}n-\frac{2}{3}=0,
\end{equation}
to find that $n$ is either $-1$ or $2/3$. Hence, the SIGF has the following solution 
\begin{equation}\label{eq: C34}
D\left(t\right)=At^{2/3}+Bt^{-1}\approx At^{2/3},
\end{equation}
where $A$ and $B$ are constants, to be determined by satisfying the initial conditions for $\delta_i$ and $\mathbf v_i$ at $z=z_0$. In the rest of the derivation, we ignore the decaying mode of the SIGF, and assume $D\propto t^{2/3}$.

Next, we solve for $E$, which obeys Eq.~\eqref{eq: C27}. During matter domination, this equation becomes
\begin{equation}\label{eq: C35}
\ddot{E}+\frac{4}{3}t^{-1}\dot{E}-\frac{2}{3}t^{-2}E=-\frac{2}{3}A^{2}t^{-2/3}.
\end{equation}
Due to zero initial conditions for $S^{(2)}$, the homogeneous solution for this equation is zero. We guess the following particular solution $E\left(t\right)=cA^{2}t^{m}$. This brings us to
\begin{equation}\label{eq: C36}
m\left(m-1\right)ct^{m-2}+\frac{4}{3}mct^{m-2}-\frac{2}{3}ct^{m-2}=-\frac{2}{3}t^{-2/3}.
\end{equation}
For this equation to hold at any time $t$, we see that we must have $m=4/3$, and thus $E=cA^{2}t^{4/3}=cD^{2}$. In order to find the constant $c$, we plug $m=4/3$ in Eq.~\eqref{eq: C36} and find $c=-3/7$. Therefore, we conclude
\begin{equation}\label{eq: C37}
E\left(z\right)=-\frac{3}{7}D^{2}\left(z\right).
\end{equation}
Note that by relaxing the assumption of $\Omega_{m}=1$, we will get a different $n$ in Eq.~\eqref{eq: C33}, which would lead to a different $m$ in Eq.~\eqref{eq: C36} and consequently a different $c$. However, Ref.~\cite{Bouchet:1994xp} showed that for a flat universe
\begin{equation}\label{eq: C38}
E\left(z\right)\approx-\frac{3}{7}\Omega_{m}^{-1/143}D^{2}\left(z\right)\approx-\frac{3}{7}D^{2}\left(z\right),
\end{equation}
where the first equality holds to a percent level.

\section{Temperature fluctuations in the dark ages}\label{sec: Temperature fluctuations in the dark ages}

In Ref.~\cite{Flitter:2023rzv} we discussed the effect of kinetic temperature fluctuations on the brightness temperature fluctuations during the dark ages. Using linear perturbation theory, we showed that the two kinds of fluctuations are approximately related in Fourier space via 
\begin{equation}\label{eq: D1}
\delta_{21,\mathrm{iso}}\left(k,z\right)\approx\frac{2\bar x_\mathrm{CMB}+\bar x_\mathrm{coll}}{\bar x_\mathrm{CMB}+\bar x_\mathrm{coll}}\delta_b\left(k,z\right)+\left[\frac{\bar x_\mathrm{CMB}}{\bar x_\mathrm{CMB}+\bar x_\mathrm{coll}}\frac{\partial\ln\bar\kappa_{1-0}^\mathrm{HH}}{\partial\ln\bar T_k}-\frac{1}{1-\bar T_k/T_\gamma}\right]\delta_T\left(k,z\right),
\end{equation}
where $\kappa_{1-0}^\mathrm{HH}$ is the collision rate (in units of $\mathrm{cm^3/sec}$) of hydrogen atoms with themselves. Here, $\delta_T\left(k,z\right)$ and $\delta_{21,\mathrm{iso}}\left(k,z\right)$ are the Fourier transform of the fractional fluctuation in $T_k\left(\mathbf x,z\right)$ and $T_{21}\left(\mathbf x,z\right)$, respectively, and the latter does not contain fluctuations sourced from peculiar velocity (see Eq.~\ref{eq: 2}). At the dark ages, the fluctuations $\delta_T$ are sourced by the perturbations in the baryon density field $\delta_b$, and thus it is useful to define the so called ``adiabatic index",
\begin{equation}\label{eq: D2}
c_T\left(k,z\right)\equiv\frac{\delta_T\left(k,z\right)}{\delta_b\left(k,z\right)}.
\end{equation}
The reason for this name is because for small $\delta_b\left(k,z\right)$ we can write $1+\delta_T\left(k,z\right)=\left[1+\delta_b\left(k,z\right)\right]^{c_T\left(k,z\right)}$.

\begin{SCfigure}
\includegraphics[width=0.5\columnwidth]{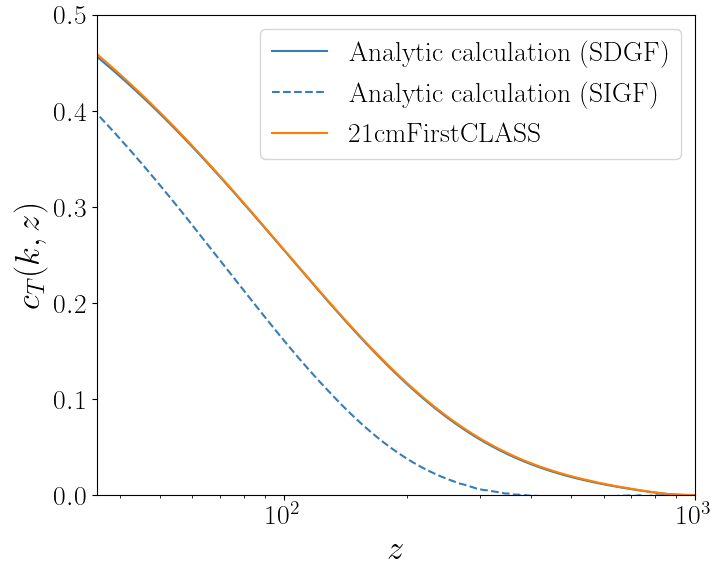}
\caption[0mm,5mm]{The scale-dependent adiabatic index for $k=0.5\,\mathrm{Mpc}^{-1}$ as a function of redshift during the dark ages. The blue curves show the analytic calculation, based on the equations found in Ref.~\cite{Flitter:2023rzv}, while the orange curve shows the numerical output from {\tt 21cmFirstCLASS}, see text for details. For the dashed blue curve, we have taken the crude approximation $\mathcal D_b\left(k,z\right)\approx D\left(z\right)$, yielding an inconsistent adiabatic index.}
\label{fig: 11}
\end{SCfigure}

This adiabatic index can be calculated analytically by solving simultaneously the linearly perturbed differential equations for $T_k$ and $x_e$. We show in Fig.~\ref{fig: 11} the analytical solution for $c_T\left(k,z\right)$ as a function of redshift for $k=0.5\,\mathrm{Mpc}^{-1}$. The blue dashed curve displays the scale-independent $c_T\left(z\right)$ which is obtained under the very crude approximation $\mathcal D_b\left(k,z\right)\approx D\left(z\right)$, namely $\delta_b\left(k,z\right)$ evolves in a scale-independent manner. In the solid blue curve we have relaxed this assumption and used the exact $\mathcal D_b\left(k,z\right)$, as defined in Eq.~\eqref{eq: 14}. For both curves, we assumed a recombination rate from the {\tt RECFAST} code~\cite{Seager:1999bc, Seager:1999km}. Evidently, the scale-dependent $c_T\left(k,z\right)$ curve reaches higher values than the scale-independent curve.

To verify the consistency of the calculations of {\tt 21cmFirstCLASS} at the dark ages we also show in Fig.~\ref{fig: 11} the output of the code for $c_T\left(k=0.5\,\mathrm{Mpc}^{-1},z\right)$. This output was achieved by computing the $c_T\left(\mathbf k,z_i\right)$ coeval box (in Fourier space) at every redshift iteration $z_i$. Because of the scale-dependence, this box is not homogeneous, and so we need to select the appropriate cells that correspond to a desired wavenumber $k_0$ (in Fig.~\ref{fig: 11}, $k_0=0.5\,\mathrm{Mpc}^{-1}$). We do this selection process using the following criterion,
\begin{equation}\label{eq: D3}
k_0-\Delta k/2 < |\mathbf k| < k_0+\Delta k/2,
\end{equation}
where $\Delta k=2\pi/L_\mathrm{box}$ and $L_\mathrm{box}$ is the box size of the simulation. Once we have the appropriate cells from the $c_T\left(\mathbf k,z_i\right)$ box, we take their median, depicted by the orange curve in Fig.~\ref{fig: 11}. There is an excellent agreement between the analytical calculation and the numerical calculation from {\tt 21cmFirstCLASS}, even though in the former we have assumed the recombination rate from {\tt RECFAST}, while the latter works with {\tt HYREC}~\cite{Ali-Haimoud:2010hou, Lee:2020obi}.

\newpage

\twocolumngrid

\bibliography{Does_it_matter.bib}

\end{document}